\documentclass[aps,prb,twocolumn,superscriptaddress]{revtex4-2}
\usepackage[colorlinks=true,allcolors=blue]{hyperref}
\usepackage{newtxtext,newtxmath}         
\usepackage{graphicx} 
\usepackage{color}
\usepackage{bm}
 \usepackage{braket}
\usepackage{xcolor}   
\usepackage{dsfont}   

\newcommand{\nc}{\newcommand}
\nc{\be}{\begin{eqnarray}}
\nc{\ee}{\end{eqnarray}}
\nc{\bea}{\begin{eqnarray}}
\nc{\eea}{\end{eqnarray}}
\nc{\bean}{\begin{eqnarray*}}
\nc{\eean}{\end{eqnarray*}}
\nc{\mb}{\mbox}
\nc{\rnc}{\renewcommand}
\nc{\vk}{{\bm k}}
\nc{\vx}{\mb{\bf x}}
\nc{\br}{\mb{\bf r}}
\nc{\bv}{\mb{\bf v}}
\nc{\bp}{\mb{\bf p}}
\nc{\ve}{\mb{\bf e}}
\nc{\vz}{\hat {\mb{\bf z}}}
\nc{\vp}{\mb{\boldmath$p$}}
\nc{\vb}{\mb{\boldmath$b$}}
\nc{\rr}{\mb{\boldmath$r$}}
\nc{\vR}{\mb{\boldmath$R$}}
\nc{\vj}{\mb{\boldmath$j$}}
\nc{\vg}{\mb{\boldmath$g$}}
\nc{\vm}{\mb{\boldmath$m$}}
\nc{\vd}{\mb{\boldmath$d$}}
\nc{\hd}{\mb{\boldmath$\hat{d}$}} 
\nc{\vD}{\mb{\boldmath$D$}}
\nc{\vF}{\mb{\boldmath$F$}}
\nc{\vG}{\mb{\boldmath$G$}}
\nc{\vI}{\mb{\boldmath$I$}}
\nc{\vW}{\mb{\boldmath$W$}}
\nc{\x}{\mb{\boldmath$x$}}
\nc{\A}{\mb{\boldmath$A$}}
\nc{\va}{\mb{\boldmath$a$}}
\nc{\vq}{\mb{\boldmath$q$}}
\nc{\vn}{\mb{\boldmath$n$}}
\nc{\vJ}{\mb{\boldmath$J$}}
\nc{\vS}{\mb{\boldmath$S$}}
\nc{\vs}{\mb{\boldmath$\sigma$}}
\nc{\vE}{\mb{\boldmath$E$}}
\nc{\vB}{\mb{\boldmath$B$}}
\nc{\vM}{\mb{\boldmath$M$}}
\nc{\vL}{\mb{\boldmath$L$}}
\nc{\vpsi}{\mb{\boldmath$\psi$}}
\nc{\vphi}{\mb{\boldmath$\varphi$}}
\nc{\Vphi}{\mb{\boldmath$\phi$}}
\nc{\Vomega}{\mb{\boldmath$\Omega$}}
\nc{\ipsi}{\it{\Psi}}
\nc{\vepsilon}{\mb{\boldmath$\epsilon$}}
\nc{\valpha}{\mb{\boldmath$\alpha$}}
\nc{\vgamma}{\mb{\boldmath$\gamma$}}
\nc{\vomega}{\mb{\boldmath$\omega$}}
\nc{\vmu}{\mb{\boldmath$\mu$}}
\nc{\vt}{\mb{\boldmath$\tau$}}
\nc{\vT}{\mb{\boldmath$T$}}
\nc{\vpi}{\mb{\boldmath$\pi$}}
\nc{\nab}{\bm{\nabla}}
\nc{\ov}{\overline}
\nc{\cdott}{\!\cdot\!}
\nc{\cdottt}{\!\!\cdot\!}
\nc{\LL}{\Big{\langle}}
\nc{\RR}{\Big{\rangle}}
\nc{\LR}{\Bigm{|}}
\nc{\vP}{\mb{\boldmath$P$}}
\nc{\nnn}{\nonumber\\}
\nc{\ltsim}{\protect\raisebox{-0.5ex}{$\:\stackrel{\textstyle <}{\sim}\:$}}
\nc{\gtsim}{\protect\raisebox{-0.5ex}{$\:\stackrel{\textstyle >}{\sim}\:$}} 
\nc{\ltsimscript}{\protect\raisebox{-0.5ex}{$\stackrel{\scriptstyle <}{\sim}$}} 
\nc{\gtsimscript}{\protect\raisebox{-0.5ex}{$\stackrel{\scriptstyle >}{\sim}$}} 
\nc{\psibar}{\overline{\psi}}
\nc{\cbar}{\overline{c}}
\nc{\intx}{\int d^4x}
\nc{\inty}{\int d^4y}
\nc{\intk}{\int \frac{d^4k}{(2\pi)^4}}
\nc{\Mhat}{\hat{\bm M}}
\nc{\tH}{t_\text{\scriptsize{H}}}
\nc{\EF}{E_\text{\scriptsize{F}}}
\nc{\chiND}{\chi^{(2)}_{j_s, \text{\scriptsize{ND}}}}
\nc{\chiD}{\chi^{(2)}_{j_s, \text{\scriptsize{D}}}}
\let\vec\boldsymbol 

\begin{document}

\title{Theory of charge and spin pumping in atomic-scale spiral magnets}

\author{Daichi Kurebayashi \href{https://orcid.org/0000-0002-5117-3791}{\includegraphics[height=0.75em]{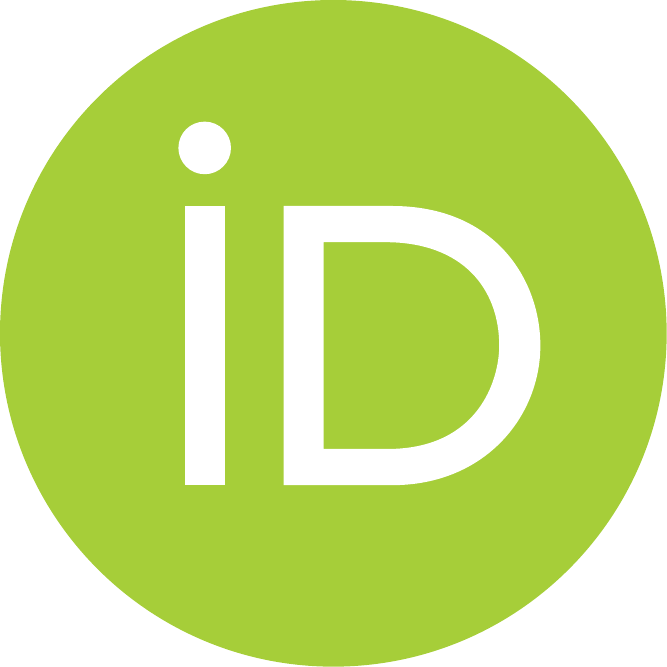}}}
\email{d.kurebayashi@unsw.edu.au}
\affiliation{RIKEN Center for Emergent Matter Science (CEMS), Wako, 351-0198, Japan}
\affiliation{School of Physics, The University of New South Wales, Sydney 2052, Australia}

\author{Yizhou Liu \href{https://orcid.org/0000-0002-0760-4179}{\includegraphics[height=0.75em]{orcid.pdf}}}
\affiliation{RIKEN Center for Emergent Matter Science (CEMS), Wako, 351-0198, Japan}

\author{Jan Masell \href{https://orcid.org/0000-0002-9951-4452}{\includegraphics[height=0.75em]{orcid.pdf}}}
\affiliation{RIKEN Center for Emergent Matter Science (CEMS), Wako, 351-0198, Japan}
\affiliation{Institute of Theoretical Solid State Physics, Karlsruhe Institute of Technology (KIT), 76049 Karlsruhe, Germany}

\author{Naoto Nagaosa \href{https://orcid.org/0000-0001-7924-6000}{\includegraphics[height=0.75em]{orcid.pdf}}}
\affiliation{RIKEN Center for Emergent Matter Science (CEMS), Wako, 351-0198, Japan}
\affiliation{Department of Applied Physics, University of Tokyo, Tokyo 113-8656, Japan}

\date{\today}

\begin{abstract}
An Archimedean screw is a classical pump that exploits the equivalence of rotation and translation in helices.
Similarly, a spin spiral texture can pump charge and spin by rotating at a frequency $\omega$.
In the present paper, we study these pumping phenomena within a microscopic quantum model by both perturbation theory and numerical simulations.
Inside the spiral region, the spin polarization and charge current are linear in $\omega$ whereas the spin current is $\omega^2$ for small $\omega$.
We find that the charge current is related to the mixed momentum-phason Berry phase, which can be viewed as a novel approximate realization of a Thouless pump. 
It is nearly quantized in spirals with short pitch $\lambda$ but decays with $\lambda^{-1}$ for longer pitches, unlike true Thouless pumps or Archimedean screws.
Moreover, we study the onset of non-adiabaticity (large $\omega$), the impact of attached non-magnetic or magnetic contacts, and the real-time evolution of the transport observables. 
Finally, we analyze the effects of disorders which, surprisingly, might enhance the spin current but suppress the charge current.
\end{abstract}

\maketitle


\section{Introduction}
\label{sec:introduction}

The concept of adiabatic charge pumping, as proposed by Thouless~\cite{Thouless1983,Niu1984}, is a fundamental topological phenomenon where a periodic change of parameters characterizing the Hamiltonian leads to a real-space shift of charge.
Experimentally, Thouless charge pumps have been realized in quantum dots~\cite{Kouwenhoven1991,Switkes1999} and cold atom systems~\cite{Lohse2016,Nakajima2016}.
In systems with a gapped spectrum, this phenomenon is described by the quantum mechanical Berry phase~\cite{Berry1984}.
In particular, the electronic states in solids are characterized by the Berry phase once the time-reversal $\cal{T}$ and/or the inversion symmetry $\cal{I}$ are broken.
For example, the electric polarization in ferroelectric insulators can be regarded as a fraction of the Thouless charge pumping associated with the displacement of the atoms from their centrosymmetric positions~\cite{KingSmith1993}.
However, the Thouless charge pump has never been realized in a macroscopic bulk sample.

\begin{figure}
\centering{
    \includegraphics[width=\columnwidth]{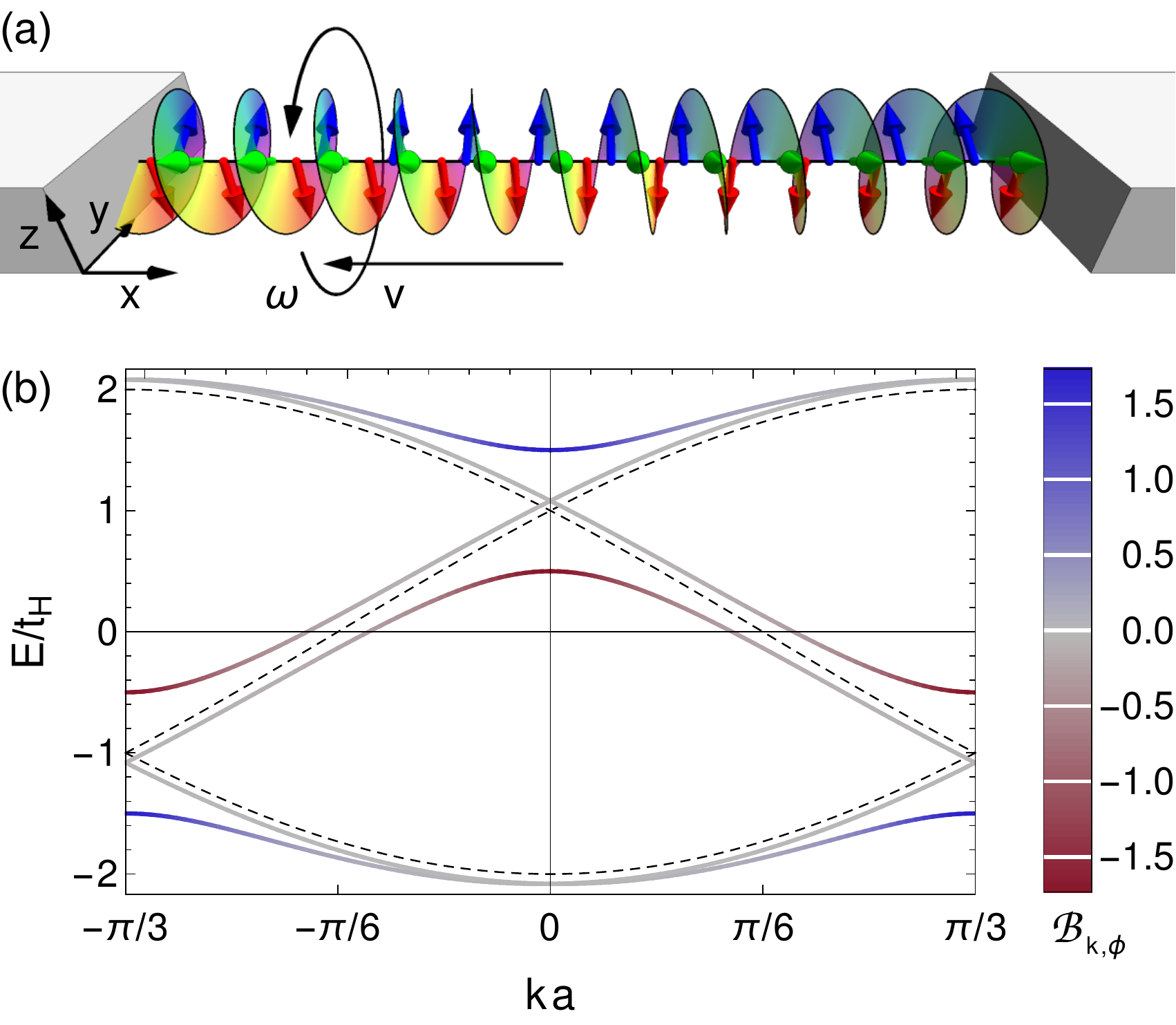}
    \caption{
    Electronic band structure $E(k)$ and sketch of the considered setup, both for $\lambda/a=3$.
    (a) Visualization of spins in a right-handed atomic-scale spiral (colored arrows) with attached leads (gray boxes).
    Rotating at a frequency $\omega$ is equivalent to translating the indicated (continuous) screw at a velocity $v=-\lambda\omega/(2\pi)$ in the direction of the  $\vec{Q}$-vector.
    (b) Dispersion $E(k)$ of a periodic system with $J/\tH = 0.5$ (colored). 
    The color encodes the momentum-phason Berry curvature $\mathcal{B}_{k,\phi}$, defined in Eq.~\eqref{eq:Berry}. 
    Dashed black lines indicate the dispersion in the absence of the spiral ($J=0$).
    } 
    \label{fig1}
}
\end{figure}

In magnets, the time-reversal symmetry $\cal{T}$ is inherently broken.
The emergent electromagnetic field which stems from the Berry phase associated with non-collinear spin textures has been studied extensively~\cite{Taguchi2001,Nagaosa2013,Schulz2012,Kanazawa2011}.
In particular, a spin spiral configuration as shown in Fig.~\ref{fig1} breaks both $\cal{T}$ and $\cal{I}$, simultaneously. 
This leads to a variety of phenomena allowed by this low symmetry, e.g., emergent inductance~\cite{Nagaosa2019,Yokouchi2020}, nonreciprocal magneto-optical effect~\cite{Tokura2014}, or nonreciprocal charge transport~\cite{Rikken2001,Jiang2020} and spin transport~\cite{Okamura2019}.
At the same time, however, spin spirals retain a peculiar symmetry:
The rotation $\mathcal{R}_{\vec{\hat{n}}}(\phi)$ of spins in the spiral plane $\vec{\hat{n}}$ is equivalent to a translation $\mathcal{T}_{\vec{Q}}(x)$ along its $\vec{Q}$-vector.
This symmetry is illustrated in Fig.~\ref{fig1} for a helical screw, i.e., where $\vec{\hat n} = \vec{Q}$.
Thus, while a rotating spin spiral naturally pumps spin into attached leads~\cite{Tserkovnyak2005a,Tserkovnyak2005b}, the simultaneously activated translational mode might also pump electric charge~\cite{Tserkovnyak2005b,delSer2021}, similar to a classical Archimedean screw pumping fluids~\cite{Vitruvius100BC,Diodorus100BC,Athenaeus200,Rorres2000}.
However, when considering a discrete lattice of atoms, the translational symmetry is broken which was neglected in previous studies.

In the present paper, we theoretically study the charge and spin pumping in a one-dimensional spiral magnet without spin-orbit interaction.
We consider systems with either
(i) periodic boundary conditions or
(ii) metallic contacts attached to both ends, see Fig.~\ref{fig1}(a).
For case (i), using perturbation theory, we derive the spin polarization $\vec{s}$, the electric current $j_e$ and spin current $\vec{j}_s$ along the $\vec{Q}$-vector and discuss their respective dependence on the driving frequency $\omega$, the spiral wavelength $\lambda$, and the Fermi energy $\EF$.
In particular, we reveal a connection between the mixed momentum-phason space Berry curvature $\mathcal{B}_{k,\phi}$ and the pumped electric current $j_e$, reminiscent of Thouless charge pumping in gapped systems.
However, half of the electronic bands remain gapless, see Fig.~\ref{fig1}(b), hence, in contrast to true Thouless pumping, the current $j_e$ is only nearly quantized in the limit of short wavelengths $\lambda/a \lesssim 10$.
Moreover, for case (ii), we numerically study both non-magnetic and magnetic leads.
Different combinations of attached leads turn out to show a very different behavior, including spin current diodes or bi-directional control via the contact magnetization.
Finally, we also study the effect of non-magnetic impurities which localize the electronic states, resulting in suppression of the electric current and, surprisingly, enhancement of the spin current, depending on the Fermi energy.

The paper is organized as follows.
In Sec.~\ref{sec:model}, we introduce the model Hamiltonian describing electrons coupled to the spin spiral and discuss its symmetries in Sec.~\ref{sec:symmetries}.
We begin the analysis by presenting the perturbation theory and its results in Sec.~\ref{sec:perturbationtheory}, including the connection to Berry curvature in Sec.~\ref{sec:berry}.
Next, we switch to non-perturbative simulation results in Sec.~\ref{sec:numerics} which focus on (i) periodic boundary conditions and open systems with non-magnetic contacts in Sec.~\ref{sec:OBCnonmagnetic} and (ii) magnetic contacts in Sec.~\ref{sec:OBCmagnetic}.
We further discuss the effect of disorder in Sec.~\ref{sec:disorder}.
Finally, we conclude with a discussion of the results and potential experimental realizations in Sec.~\ref{sec:discussion}.
Additional details are provided in the Appendix.


\section{The Model}
\label{sec:model}

In the following, we first introduce the basic model which is the basis of this study.
Next, we briefly discuss symmetries of this model and their implications for the universality of our results and, finally, we shortly review the electronic band structure of the static model.

\subsection{Magnetization, electrons, observables}
\label{sec:modeldetails}
We are interested in the electronic response to rotations/translations of a magnetic spiral; see Fig.~\ref{fig1}(a) for a schematic sketch of the setup.
To this point, for simplicity, we neglect the influence of conduction electrons on the magnetization dynamics in our one-dimensional toy model.
Instead, we postulate that the magnetization is described by
\begin{equation}
\vec{M}(x,t) = \vec{\hat e}_1 \cos\theta(x,t) + \vec{\hat e}_2 \sin\theta(x,t) \,\,,
\label{eq:M}
\end{equation}
where $\theta(x,t) = Q x + \phi(t)$ is the phase of the spiral.
The normalized vectors $\vec{\hat e}_1$ and $\vec{\hat e}_2$ define the spiral plane, i.e., the $y$-$z$-plane in Fig.~\ref{fig1}(a).
The wavelength $\lambda$ enters via $Q = 2\pi\eta/\lambda$, where $\eta$ distinguishes right-handed ($\eta=1$) and left-handed spirals ($\eta=-1$).
Within this simplified approach, all magnetization dynamics are reduced to the dynamics of the time-dependent phason $\phi(t)$:
As indicated in Fig.~\ref{fig1}(a), the phason dynamics $\phi(t)$ can be interpreted as both, a rotation with an angular velocity $\omega$ or a translation with velocity $v=-\eta\lambda\omega/(2\pi)$.
Our discussion of results in Sec.~\ref{sec:results} is based on the right-handed helix in Fig.~\ref{fig1}(a).
However, other the results for different spiral planes or handedness are related by symmetry, see Sec.~\ref{sec:symmetries}. 

The electronic system with its full dynamics is described by a standard tight-binding model
\begin{equation}
 \mathcal{H} = \sum_j - \,\tH \,\left(\vec{c}_{j+1}^\dagger \vec{c}_{j} + h.c. \right)\, - J \,\vec{c}_j^\dagger (\bm{\sigma} \!\cdot\! \bm{M})\, \vec{c}_j \,\,.
\label{eq:H} 
\end{equation}
We use the spinor notation $\bm c_{j} = \left( c_{j,\uparrow}, c_{j,\downarrow} \right)^T$ for the up/down-spin electron annihilation operators $c_{j,\uparrow/\downarrow}$ at site $j$ and $\bm\sigma$ is a 3-vector which contains the Pauli matrices.
The transfer integral $\tH$ describes hopping of electrons between adjacent sites and the exchange constant $J$ couples the electrons to the magnetization.

The rotating spiral induced dynamics in the electronic system.
In order to quantify these dynamics, we evaluate the spin polarization $\vec s$, the charge current $j_e$, and the spin current $\vec{j}_s$.
Note that the vector notation of $\vec s$ and $\vec{j}_s$ refers to the spin-components.
The real-space direction of the currents $j_e$ and $\vec{j}_s$ is given by the spiral's $\vec{Q}$-vector and, hence, we neglect it in the following.
For the observables, the local quantum mechanical operators for site $i$ and spin-component $\alpha=x,y,z$ are given by
\begin{align}
 \hat{s}_{\alpha,i} &= \bm c_j^\dagger \sigma_\alpha \bm c_j^{} \,, \label{eq:sx}\\
 \hat{j}_{e,i} &= - q\, i \,\tH \left(\bm c_{j+1}^\dagger \bm c_{j}^{} - \bm c_{j}^\dagger \bm c_{j+1}^{}\right) \,, \label{eq:je}\\
 \hat{j}_{s,\alpha,i} &= - i \,\tH \left(\bm c_{j+1}^\dagger \sigma_\alpha \bm c_{j}^{} - \bm c_{j}^\dagger \sigma_\alpha \bm c_{j+1}^{}\right) \,\,.  \label{eq:js}
\end{align}
with $q=-e<0$ the electron charge.
In the following, the average $\langle ... \rangle$ denotes the temporal average, if not stated otherwise, and the site index $i$ is dropped for convenience. 
Moreover, for the helical screw in Fig.~\ref{fig1} that we consider during most of this study, only the spin polarization $\langle s_x \rangle$ and spin current $\langle j_{s,x} \rangle$ can be non-zero by symmetry.


\subsection{Symmetries and generalization}
\label{sec:symmetries}

As motivated in the introduction, even though the non-magnetic part of the tight-binding Hamiltonian $\mathcal{H}$ in Eq.~\eqref{eq:H} is highly symmetric, the magnetic texture breaks both time-reversal $\cal{T}$ and inversion symmetry $\cal{I}$.
However, some symmetries remain intact or can be restored by proper transformation of the magnetic spiral.
When combined with the symmetries of the observables in Eqs.~\eqref{eq:sx}-\eqref{eq:js}, they reveal characteristic features or can be used to transfer our results for the specific system in Fig.~\ref{fig1} to other spirals.
The details are discussed in the following with a summary provided in Tab.~\ref{tab:symmetries}.

\begin{table}[t]
    \begin{center}
    \begin{tabular*}{0.93\columnwidth}{ c|c| *{4}{c} }  
        \kern 2mm transformation \kern 2mm& \kern 2mm symmetry \kern 2mm & ${v}$ & $\vec{s}$ & $j_e$ & $\vec{j}_s$  \\ 
        \hline\hline
        \kern -3mm $\phantom{\hat{\mathcal{R}}} \vec{e}_3 \to \hat{\mathcal{R}}\vec{e}_3$ & SU(2) & \kern 2mm --- \kern 2mm & $\hat{\mathcal{R}}\vec{s}$ & \kern 2mm --- \kern 2mm & $\hat{\mathcal{R}}\vec{j}_s$ \\ 
        \hline
        \kern -3mm $\phantom{-} \eta\to-\eta$ & $\mathcal{I}$ & $-{v}$ & \kern 2mm --- \kern 2mm & $-{j}_e$ & $-\vec{j}_s$ \\ 
        \hline
        \kern -3mm $\phantom{-} \omega\to-\omega$ & $\mathcal{T}$ & $-{v}$ & $-\vec{s}$ & $-{j}_e$ & \kern 2mm --- \kern 2mm \\ 
        \hline
        \kern -3mm $\phantom{-} \EF\to-\EF$ & $\Xi$ & \kern 2mm --- \kern 2mm & \kern 2mm --- \kern 2mm & $-{j}_e$ & \kern 2mm --- \kern 2mm \\ 
        \hline
    \end{tabular*} 
    \end{center}
    \caption{
    Overview of transformations, related symmetries, and their effects on the spiral velocity ${v}$, spin accumulation $\vec{s}$, charge current ${j}_e$, and spin current $\vec{j}_s$, assuming a constant angular velocity $\omega$ of the spiral.
    The transformations are a rotation $\hat{\mathcal{R}}$ of the normal vector of the spiral plane $\vec{\hat e}_3$ in spin-space, inverting the handedness $\eta$, inverting the rotation frequency $\omega$, and inverting the Fermi energy $\EF$.
    A horizontal bar indicates that the respective quantity is invariant under a transformation.
    Details are given in the main text.
    }
    \label{tab:symmetries}
\end{table}

\textit{U(1) phase ---} 
A time-independent spiral phase $\phi$ of Hamiltonian $\mathcal{H}$ is absorbed by the spin-dependent U(1) gauge transformation, $c_{j, \uparrow (\downarrow)} \rightarrow c_{j, \uparrow (\downarrow)} e^{- \sigma i\phi/2}$ where $\sigma = + (-) 1$ for up (down) spin.
Meanwhile, the observables given in Eqs.~\eqref{eq:sx}-\eqref{eq:js} are invariant under the gauge transformation.
This implies that the time-averaged observables are independent of the phason $\phi(t)$ itself and can only depend on derivatives thereof.
Moreover, the only symmetry-allowed direction for the spin accumulation and spin current is perpendicular to the spiral plane.

\textit{SU(2) phase ---} 
The aforementioned U(1) symmetry is generalized to an SU(2) symmetry, i.e., rotations of all spins with respect to any common axis leave the electronic part of $\mathcal{H}$ invariant.
Similar to the U(1) case, the magnetic texture needs to be rotated which may not only involve the phason $\phi(t)$ but the entire spiral plane defined by its normal vector $\vec{\hat e}_3 = \vec{\hat e}_1 \times \vec{\hat e}_2$.
While the charge current ${j}_e$ is also SU(2) symmetric, the spin accumulation $\vec{s}$ and current $\vec{j}_s$ are not.
For both $\vec{s}$ and $\vec{j}_s$, the only non-vanishing spin component is given by the rotated normal vector $\vec{\hat e}_3 = \hat{\mathcal{R}}\vec{\hat e}_x$.
The results presented in the following for the righthanded helix in Fig.~\ref{fig1} with $\vec{\hat e}_3 = \vec{\hat e}_x$ thus directly apply also to righthanded cycloids or arbitrary spiral planes with the rotated spiral plane $\vec{\hat e}_3$.

\textit{Inversion $\mathcal{I}$ ---} 
Upon inversion $\mathcal{I}$, the handedness $\eta$ of the magnetic spiral is inverted.
Simultaneously, the signs of both currents ${j}_e$ and $\vec{j}_s$ are flipped as can be seen directly from in Eqs.~\eqref{eq:je} and \eqref{eq:js} where exchanging site indices leads to an additional minus sign.
The sign of the spin accumulation $\vec{s}$ (and the hopping term in $\mathcal{H}$) remain unaffected.

\textit{Time reversal $\mathcal{T}$ ---} 
Changing the sign of the rotation frequency $\omega \to -\omega$ (or drift velocity $v$) is obtained by rotating the entire setup which corresponds to a combined time reversal and a phase shift of $\pi$.
While any magnetic texture breaks $\mathcal{T}$ symmetry with $\mathcal{T}(\vec{M})=-\vec{M}$ (or $\mathcal{T}(\vec{c}_{i, \sigma})=-\vec{c}_{i, -\sigma}$), in a spiral this is equivalent to a phase shift of $\pi$. 
Thus, it can be gauged away by exploiting the U(1) symmetry of the electronic system but switches the sign of the spin-dependent observables.
Therefore, the symmetry $\mathcal{T}$ switches the signs of the spin accumulation $\vec{s}$ and the current ${j}_e$, but the doubled sign change in the spin current $\vec{j}_s$ cancels.

\textit{Particle-hole symmetry $\Xi$ ---} 
The composite symmetry $\Xi = \Gamma \mathcal{T}$, composed of the chiral symmetry $\Gamma(\vec{c}_{i, \sigma}) = (-1)^i \vec{c}_{i, \sigma}$ and time reversal symmetry $\mathcal{T}$, is anti-unitary with $\Xi^2 = -1$.
It transforms the Hamiltonian as $\Xi^{-1} \mathcal{H} \Xi = - \mathcal{H}$.
Thus, $\Xi$ acts as particle-hole symmetry, i.e., it changes the sign of the Fermi energy $\EF$.
While the spin accumulation $\vec{s}$ and spin current $\vec{j}_s$ are invariant under $\Xi$, the current ${j}_e$ changes its sign.


\subsection{Helical and spin-density wave channels}
\label{sec:channels}

Before we study the response properties of the dynamical system, let us briefly summarize the electronic properties of the static system.

The electronic band structure for $\lambda/a=3$ is shown in Fig.~\ref{fig1}(b).
For comparison, Fig.~\ref{fig1}(b) also includes the band structure without magnetic order where all bands are 2-fold degeneracy.
At crossing points, $E/\tH = \pm 1$, the latter system with $J=0$ realizes a 4-fold degenerate point.
In contrast, for finite $J\neq0$ one pair of bands hybridizes and opens a gap whereas the other pair of bands remains gapless.
This observation for $\lambda/a=3$ is universal.

For any $\lambda>2a$, due to the existence of spiral order, one of two degenerate spin state pairs hybridizes and opens a spin density-wave (SDW) gap of size $\Delta = 2|J|$ at momenta where the nesting condition is satisfied.
In one-dimensional systems, the SDW gaps appear at two different energies; the centers of upper and lower SDW gap are located at 
\begin{equation}
E_{\rm gap} = \pm 2\tH \cos Qa/2 \,\,, 
\end{equation}
respectively, where $a$ is the lattice constant.
The other pair of bands remains gapless.
The gapless channels inside of the SDW gaps are fully spin-polarized, so-called helical channels.
As a result, only the gapped (hybridized) SDW channels carry a high concentration of mixed momentum-phason Berry curvature $\mathcal{B}_{k,\phi}$, which is defined further below in Eq.~\eqref{eq:Berry}.
In turn, the in-gap helical channels do not carry Berry curvature.
Nonetheless, they play an important role and are responsible for the deviation from quantized Thouless/Archimedes transport, see Appendix~\ref{sec:collinearSDW} for details.


\section{Results}
\label{sec:results}
In this section, we calculate the spin polarization $\vec s$, the charge current $j_e$, and the spin current $\vec{j}_s$ induced by a rotational/sliding motion of the spin spiral. 
Starting from our model without spin-orbit coupling, Sec.~\ref{sec:modeldetails}, we derive semi-analytical perturbative results in Sec.~\ref{sec:perturbationtheory} which we compare to results of our numerical non-perturbative analysis based on the time-dependent Schr\" odinger equation in Sec.~\ref{sec:numerics}.
Based on the latter technique, we reveal the effect of magnetic contacts and disorder, respectively, in Sec.~\ref{sec:OBCmagnetic} and \ref{sec:disorder}.


\subsection{Perturbation theory with periodic boundary conditions}
\label{sec:perturbationtheory} 

As discussed in Sec.~\ref{sec:symmetries}, due to the U(1) gauge symmetry, the observables can only depend on time-derivatives of $\phi(t)$.
Therefore, the nonequilibrium physical observables $\braket{\mathcal{O}(t)}$ can be expressed in powers of the time-derivative of $\phi (t)$:
\begin{equation}
\braket{\mathcal{O}}(t) = \chi_{\mathcal{O}}^{(1)} \left(\frac{d\phi(t)}{dt}\right) + \chi_{\mathcal{O}}^{(2)}\left(\frac{d\phi(t)}{dt}\right)^2 + \cdots \,\,.
\label{eq:expectationvalue}
\end{equation}
Here, $\chi_{\mathcal{O}}^{(1)}$ and $\chi_{\mathcal{O}}^{(2)}$ are the first- and second-order susceptibilities with respect to $\phi(t)$, respectively.
Moreover, the various symmetries, summarized in Tab.~\ref{tab:symmetries}, impose further constraints which determine the leading order contributions in $\omega$ to the individual observables.
In particular, the signs of the spin polarization $\vec s$ and charge current $j_e$ switch with the sign of $\omega$, allowing for only terms with odd powers.
In turn, the spin current $\vec{j}_s$ is even in $\omega$, thus its leading order contribution stems from the second order susceptibility $\chi_{j_s}^{(2)}$.

Similarly, we can expand the Hamiltonian from Eq.~(\ref{eq:H}) with respect to the phason $\phi(t)=\omega t$ and obtain 
\begin{equation}
\mathcal{H}(t) \approx \mathcal{H}_0 
- J \phi(t) \sum_i \bm c_i^\dagger \hat{\Sigma}_{\phi}^{(1)} \bm c_i 
- \frac{J}{2} \phi^2(t) \sum_i \bm c_i^\dagger \hat{\Sigma}_{\phi}^{(2)} \bm c_i 
 \,\,,
\end{equation}
where $\mathcal{H}_0 = \mathcal{H}(t=0)$ is the initial Hamiltonian.
The operators $\hat{\Sigma}_{\phi}^{(1)}$ and $\hat{\Sigma}_{\phi}^{(2)}$ coupled to the phason are straightforwardly obtained as first and second order derives of $\vec \sigma \cdot \vec M$ with respect to $\phi$, respectively.

In the following, we discuss the resulting perturbative expressions for the observables based on a Feynman diagram picture.
Details on the calculations are provided in Appendix~\ref{sec:perturbationdetails}.


\begin{figure*}[t]
	\includegraphics[width = \textwidth]{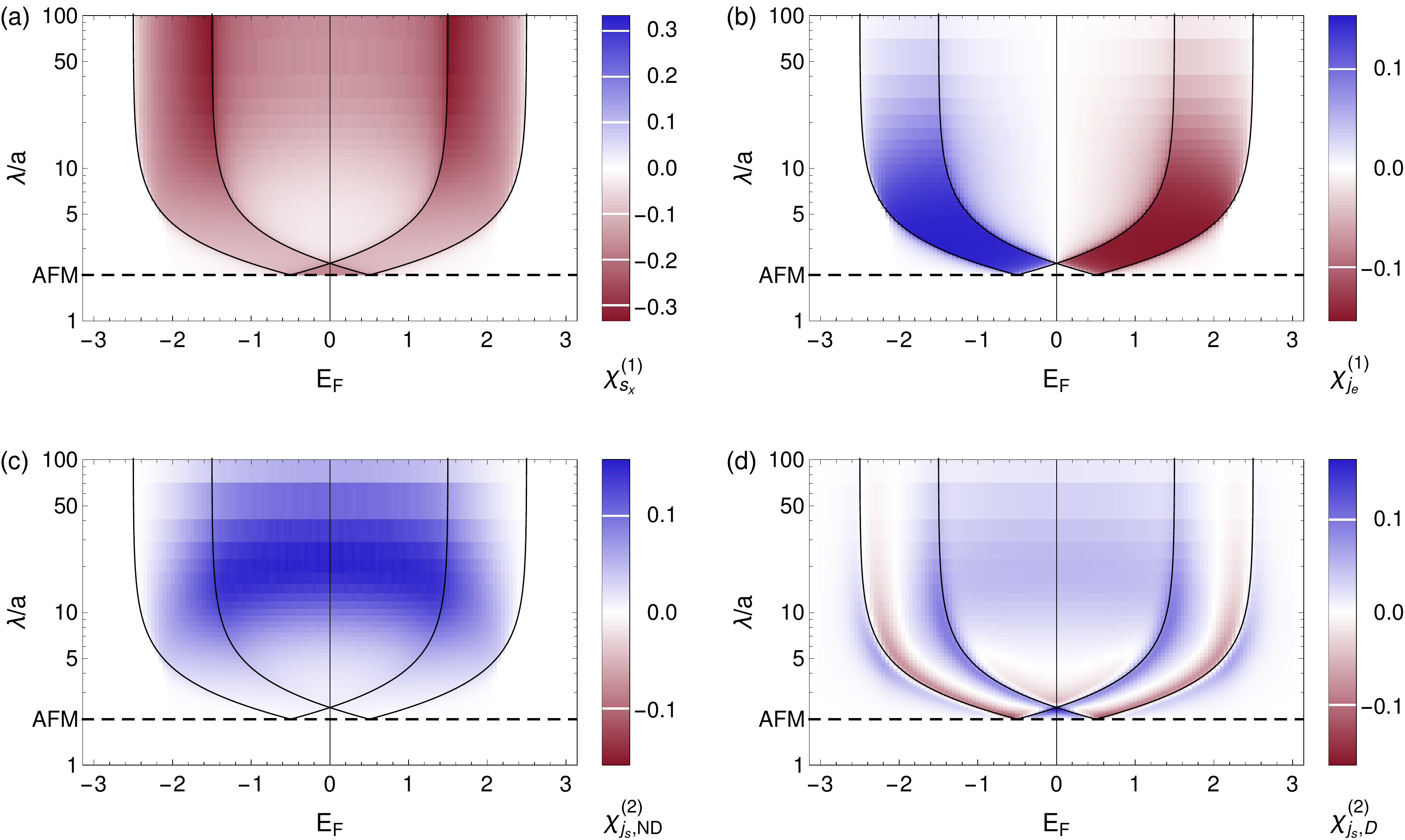}
	\caption{
	First- and second-order susceptibilities corresponding to (a) the spin polarization $s_x \propto \omega$, (b) the charge current $j_e \propto \omega$, and (c) the non-dissipative and (d) dissipative contribution to the spin current $j_{s,x} \propto \omega^2$, see Eq.~\eqref{eq:expectationvalue}.
	The susceptibility as function of the Fermi energy $\EF$ and wavelength $\lambda/a$ is encoded by color, indicated in each panel.
	The horizonal dashed line indicates $\lambda=2a$, i.e., the collinear antiferromagnetic order, below which we don't show data.
	Solid black lines indicate the edges of the SDW gap, given by $\lambda(\EF) = \pi a / \arccos\left(\pm(\EF \pm J)/(2\tH)\right)$.
	Parameters are $t_\text{\tiny{H}} = 1$, $J = 0.5$, and $\tau = 2.5$.
	}
	\label{fig2}
\end{figure*}


\subsubsection{Linear order response}
\label{sec:order1}

An approximation to linear order in $\omega$ captures the spin accumulation $s_x$ and the charge current $j_e$, with the Feynman diagrams shown as insets in Figs.~\ref{fig2}(a) and (b).
To this level, the particle-hole bubble (solid lines) only couples the observable's operator $\hat{\mathcal{O}}=\hat{s}_x$ or $\hat{j}_e$ (wiggled line) to the linear-order phason-related operator $\hat{\Sigma}_{\phi}^{(1)}$ (dashed line).
Thus, the first-order susceptibility  shown in Figs.~\ref{fig2}(a,b) reads
\begin{widetext}
\begin{equation}
\chi_{j_e}^{(1)} = 2 J 
\int_\text{BZ}\frac{d k}{2\pi}\, 
\sum_{i \neq j}\, 
\frac{f\left(\epsilon_{k,i} - \EF\right)}{\left(\epsilon_{k,i} - \epsilon_{k,j}\right)^2}
{\rm Im}\left[\langle\psi_0^{k,i}| \hat{j}_e |\psi_0^{k,j}\rangle\, 
\langle\psi_0^{k,j}| \hat{\Sigma}_{\phi}^{(1)} |\psi_0^{k,i}\rangle\, \right]
\,\,.
\label{eq:order1}
\end{equation}
\end{widetext}
In our notation, the states $\psi_0^{k,i}$ are eigenstates of the unperturbed Hamiltonian $\mathcal{H}_0$ with momentum $k$, band index $i$, and eigenenergy $\epsilon_{k,i}$.
The $k$-integral is over the entire Brillouin zone (BZ) and $f(E)=(e^{E/k_\text{\scriptsize{B}} T}+1)^ {-1}$ denotes the Fermi-Dirac distribution, here with $k_\text{B}=1$.

The linear order susceptibilities given in Eq.~\eqref{eq:order1} are shown in Fig.~\ref{fig2} for the spin accumulation, panel (a), and the charge current, panel (b), explicitly calculated for a right-handed spiral with $\tH=1$ and $J=0.5$.
The graphs show the susceptibilities as function of both the Fermi energy $\EF$ and the wavelength $\lambda$, simultaneously.
In agreement with the discussion of symmetries Sec.~\ref{sec:symmetries}, the spin accumulation $s_x$ is symmetric in $\EF$ whereas the charge current $j_e$ is antisymmetric around $\EF=0$, i.e., it depends whether the charge carriers are particle-like or hole-like.

More precisely, bound by the lower edge of the lower SDW gap and the upper edge of the upper SDW gap, the spin accumulation $s_x$ induced by the rotation of the magnetic background follows a left-hand rule and, hence, is always negative.
For larger wavelength $\lambda$, the susceptibility of the spin accumulation $s_x$ plateaus quickly at the value for the ferromagnetic limit.
Exactly in the centers of the SDW gaps, $\EF=\pm2\tH$, $\chi_{s_z}^{(1)}$ agrees well with the result of our effective continuum model, which is discussed in Appendix~\ref{sec:continuum}, namely
\begin{equation}
 \chi_{s_z}^{(1)} = - \frac{1}{2\pi a\sqrt{J \tH}} + \frac{\pi a}{2\lambda^2} \sqrt{\frac{\tH}{J^3}} \quad\text{for } \lambda \gg  \pi a\sqrt{2\tH/J}\,\,.
\end{equation}
However, note that larger values of $\chi_{s_z}^{(1)}$ are obtained closer to the inner edges of the SDW gaps, $\EF\to\pm(2 \tH- J)$.
In the opposite limit, $\lambda/a\to2$, the SDW gaps come closer and therefore the interval of Fermi energies $\EF$ with a finite response becomes considerably smaller.
Meanwhile, the maximum value decreases but it remains finite down to the antiferromagnetic limit $\lambda/a=2$ as there is no symmetry which forbids spin accumulation in the collinear antiferromagnet (AFM).

The charge current $j_e$ starts off positive for $\EF<0$ as the drift velocity $v$ of a right-handed spiral ($\eta=1$) is negative, see Fig.~\ref{fig1}, and also the charge of the electronic carriers is negative, $q=-e<0$
Moreover, the susceptibility is finite mostly within the SDW gaps but quickly decays outside as well as for increasing wavelength $\lambda\to\infty$.
Similar to the spin accumulation $s_x$, our effective continuum model describes the limit of large wavelength and additionally predicts a plateau, i.e., quantized transport, in the limit of small wavelength, see Appendix~\ref{sec:continuum}.
However, the latter has to break down in the AFM limit $\lambda/a=2$:
Cases $\lambda<2a$ are equivalent to $\lambda>2a$ with inverted handedness $\eta$.
Accordingly, the charge current is antisymmetric with respect to both $\EF=0$ and $\lambda=2a$, where it must vanish, respectively.
These limiting cases  can be summarized as
\begin{equation}
 \chi_{j_e}^{(1)} = 
 \left\{
 \begin{array}{ll}
    e a \sqrt{\tH / J}/\lambda, & \text{for } \lambda \gg  \pi a\sqrt{2\tH/J}\\
    e/2\pi, & \text{for } 2a \lesssim \lambda \ll  \pi a\sqrt{2\tH/J}\\
    0, & \text{for } \lambda = 2a
\end{array}
\right. .
\end{equation}
The maximal charge current $j_e$ is therefore expected in the plateau region with $\lambda$ just a few lattice sites $a$.

The linear response results for the charge current $j_e$ are in stark contrast to the behavior that would be expected from a classical Archimedean screw, i.e., where the current would be linear in the velocity $v=-\lambda\omega/(2\pi)$ and, thus, also in $\lambda$ \cite{delSer2021}.
In contrast, the narrow plateau region close to the AFM limit with almost quantized transport is an example of almost quantized Thouless pumping.
However, unlike charge density waves or collinear spin density waves, the spin spiral has gapless helical channels which spoil the quantization.
Our result is therefore in agreement with the naive expectation that charge pumping is absent in a fully polarized ferromagnet ($\lambda\to\infty$).
We discuss the role of the momentum-phason Berry curvature $\mathcal{B}_{k,\phi}$ in the following section.
A detailed comparison to the truly quantized Thouless pumping in a collinear spin density wave is presented in Appendix~\ref{sec:collinearSDW}.


\subsubsection{Berry curvature}
\label{sec:berry}

As mentioned already in the previous sections, the mixed momentum-phason Berry curvature $\mathcal{B}_{k,\phi}$ determines the charge current $j_e$ induced by a rotation of the spin spiral.
Using the relations
\begin{equation}
\hat{j}_e = -(-e) \frac{d \hat{\mathcal{H}}}{d k} 
\quad \text{and} \quad
\hat{\Sigma}_\phi^{(1)} = -\frac{1}{J}\frac{d \hat{\mathcal{H}}}{d \phi} \,\,,
\end{equation}
and after some algebra, we can derive an additional expression for the momentary charge current $j_e$ which reads
\begin{equation}
j_e(t) = -e \int_\text{BZ}\frac{d k}{2\pi}\, \sum_i f\left(\epsilon_{k,i} - \EF\right) \mathcal{B}_{k,\phi}^i \frac{d \phi}{d t} \,\,.
\end{equation}
Here, we have introduced the mixed momentum-phason space Berry curvature 
\begin{equation}
\mathcal{B}_{k,\phi}^i = i \left(\braket{\partial_k \psi^{k,i} | \partial_\phi \psi^{k,i}} - \braket{\partial_\phi \psi^{k,i} | \partial_k \psi^{k,i}}\right) \,\,,
\label{eq:Berry}
\end{equation}
defined in the space spanned by the momentum $k$ and the phason $\phi$.
As for Eq.~\eqref{eq:order1}, $\psi^{k,i}$ denotes an eigenstate of the Hamiltonian $\mathcal{H}$ with given $\phi$, $i$ is a band index, and $f(E)$ is the Fermi-Dirac distribution.
After averaging over a  period of the spiral motion we obtain the average charge current
\begin{equation}
\braket{j_e} = \frac{-e}{4\pi^2} \int_{\rm BZ} dq \int_0^{2\pi} \!\!\!d\phi \sum_i  f\left(\epsilon_{k,i} - \EF\right) \mathcal{B}_{k,\phi}^i \,\,.
\end{equation}
Consequently, the pumped charge is given by the sum over the Berry curvature $\mathcal{B}_{k,\phi}^i$ of all occupied states.
However, since most of the Berry curvature is concentrated at the edges of the SDW gaps, the total current pumped by a single cycle of rotation is \emph{almost} quantized to $e$ when the Fermi energy is located in the SDW gap.
Moreover, since the Berry curvature on both edges of the SDW gap cancels, there is almost no charge transport outside of the SDW gaps.

The perfect quantization is hindered due to the existence of the gapless helical states.
This apparent difference to collinear spin density waves which, in turn, do feature quantized transport, is discussed in Appendix~\ref{sec:collinearSDW}.
Most importantly, as the spectrum is not fully gapped, the Berry curvature is not confined to an increasing number of isolated bands when increasing the wavelength $\lambda$ but, instead, it smears out across the SDW gap via the gapless bands and finally cancels with the curvature from the other bands.


\subsubsection{Second order response}
\label{sec:order2}

Because of the symmetry constraints, c.f. Tab.~\ref{tab:symmetries}, the pumped spin current $j_s$ is even in the driving frequency $\omega$.
Therefore, the linear order perturbation theory in the previous Sec.~\ref{sec:order1} does not yield any contributions to $j_s$.
Instead, finite contributions only enter on the second order level, $j_s \propto \omega^2 + \mathcal{O}(\omega^4)$, via the two Feynman diagrams shown as insets in Figs.~\ref{fig2}(c) and (d).
To second order in $\omega$, similar to the linear order case, the particle-hole bubble (solid lines) couples the observable's operator $\hat{j}_s$ (wiggled line) to the second-order phason-related operator $\hat{\Sigma}_{\phi}^{(2)}$ (two joined dashed lines, panel (c)).
However, an additional contribution arises since the operator $\hat{j}_s$ may also couple twice to the linear-order phason-related operator $\hat{\Sigma}_{\phi}^{(1)}$ (two separate dashed lines, panel (d)).
In analogy to diamagnetic and paramagnetic contributions of electric conductivity, these two contributions cancel the terms explicitly depending on $\phi(t)$ to preserve the U(1) symmetry.
See Appendix~\ref{sec:perturbationdetails} for the detailed discussion.
The second-order susceptibility reads
\begin{widetext}
\begin{align}
\chi_{j_s}^{(2)} & = \chiND + \chiD, \\
\chiND &= 3J^2
\int_\text{BZ}\frac{d k}{2\pi} \,
\sum_{i \ne j} \,
\langle\psi_0^{k,i}| \hat{j}_s |\psi_0^{k,i}\rangle\, 
|\langle\psi_0^{k,i}| \hat{\Sigma}_{\phi}^{(1)} |\psi_0^{k,j}\rangle |^2\, 
\frac{f\left(\epsilon_{k,i} - \EF\right) - f\left(\epsilon_{k,j} - \EF\right)}{\left(\epsilon_{k,i} - \epsilon_{k,j}\right)^4} \,\,,
\label{eq:order2nd} \\
\chiD &= \frac{iJ^2}{8}
\int_\text{BZ}\frac{d k}{2\pi} \,
\sum_{i \ne j} \,
\langle\psi_0^{k,i}| \hat{j}_s |\psi_0^{k,i}\rangle\, 
|\langle\psi_0^{k,i}| \hat{\Sigma}_{\phi}^{(1)} |\psi_0^{k,j}\rangle |^2\, 
\left\{ \left( g^r_{q, i} \right)^2 - \left( g^a_{q, i}\right)^2 \right\} \left(g^r_{q, j} + g^a_{q, j}\right)^2\,\,,
\label{eq:order2d}
\end{align}
\end{widetext}
where we again used the notation as in Eq.~\eqref{eq:order1} and, additionally, 
$\hat{j}_s$ is the spin current operator, 
$g^{r(a)}_{q, i} = (\EF - \epsilon_{q,i} \pm \frac{i}{2\tau})$ is the retarded (advanced) Green's function of the unperturbed Hamiltonian $\mathcal{H}_0$, 
and $\tau$ is a scattering lifetime.
Equation~\eqref{eq:order2nd} emerges from contributions of all the states below the Fermi energy, known as the Fermi sea contribution, and describes a non-dissipative spin current which is almost independent of impurity scattering.
On the other hand, Eq.~\eqref{eq:order2d} contains contributions from excitations near the Fermi surface, therefore, gives dissipative spin current.
In the following, we refer the susceptibility given in Eqs.~\eqref{eq:order2nd} and~\eqref{eq:order2d} as non-dissipative and dissipative contributions, respectively.
Note that $\chiD$ is proportional to $1/\tau$ for large $\tau$, i.e., induced by the relaxation.

The non-dissipative and dissipative second order susceptibilities are shown in Fig.~\ref{fig2}(c) and (d), respectively, both as function of the Fermi energy $\EF$ and the wavelength $\lambda/a$.
The other parameters are chosen as $\tH=1$, $J=0.5$, and $\tau=2.5$. 
The plots reflect the symmetries that were already discussed in Sec.~\ref{sec:symmetries} and are summarized in Tab.~\ref{tab:symmetries}.
Moreover, as discussed in Sec.~\ref{sec:order1}, the antisymmetric behavior in the handedness $\eta$ implies that the spin current $j_s$ has to vanish in the AFM limit, $\lambda/a\to2$.

As shown in Fig.~\ref{fig2}(c), the non-dissipative second order susceptibility $\chiND$ is always positive for all Fermi energies $\EF$ and wavelengths $\lambda$.
In contrast to the spin accumulation $s_x$ and charge current $j_e$, its range is not clearly bound by the edges of the SDW gaps and looks more complex with a maximum for Fermi energies $\EF$ located between the SDW gaps and $\lambda \sim 10 a$.
In the large wavelength limit, our continuum approximation gives the asymptotic behavior in the center of the SDW gaps 
\begin{equation}
\chiND = \frac{e a}{2\lambda} \sqrt{\frac{\tH}{J^3}} \quad\text{for } \lambda \gg  \pi a\sqrt{2\tH/J}\,\,,
\end{equation}
see Appendix~\ref{sec:continuum} for details.

The dissipative second order susceptibility $\chiD$, see Fig.~\ref{fig2}(d), displays a quite different behavior.
Here, finite contributions are mostly located at the edges of the SDW gaps where the mixed momentum-phason Berry curvature $\mathcal{B}_{k,\phi}$ is concentrated.
Similar to the Berry curvature $\mathcal{B}_{k,\phi}$, the sign of the spin current $j_s$ is also opposite on the two opposite edges of each SDW gap.
Therefore, our analytical continuum approximation for the centers of the SDW gaps is meaningless for the dissipative susceptibility.
Still, we observe that in the ferromagnetic limit $\lambda\to\infty$ the expected result $j_s\to0$ is obtained.


\subsection{Numerical calculation}
\label{sec:numerics}

\begin{figure*}[t]
\centering{
    \includegraphics[width=\textwidth]{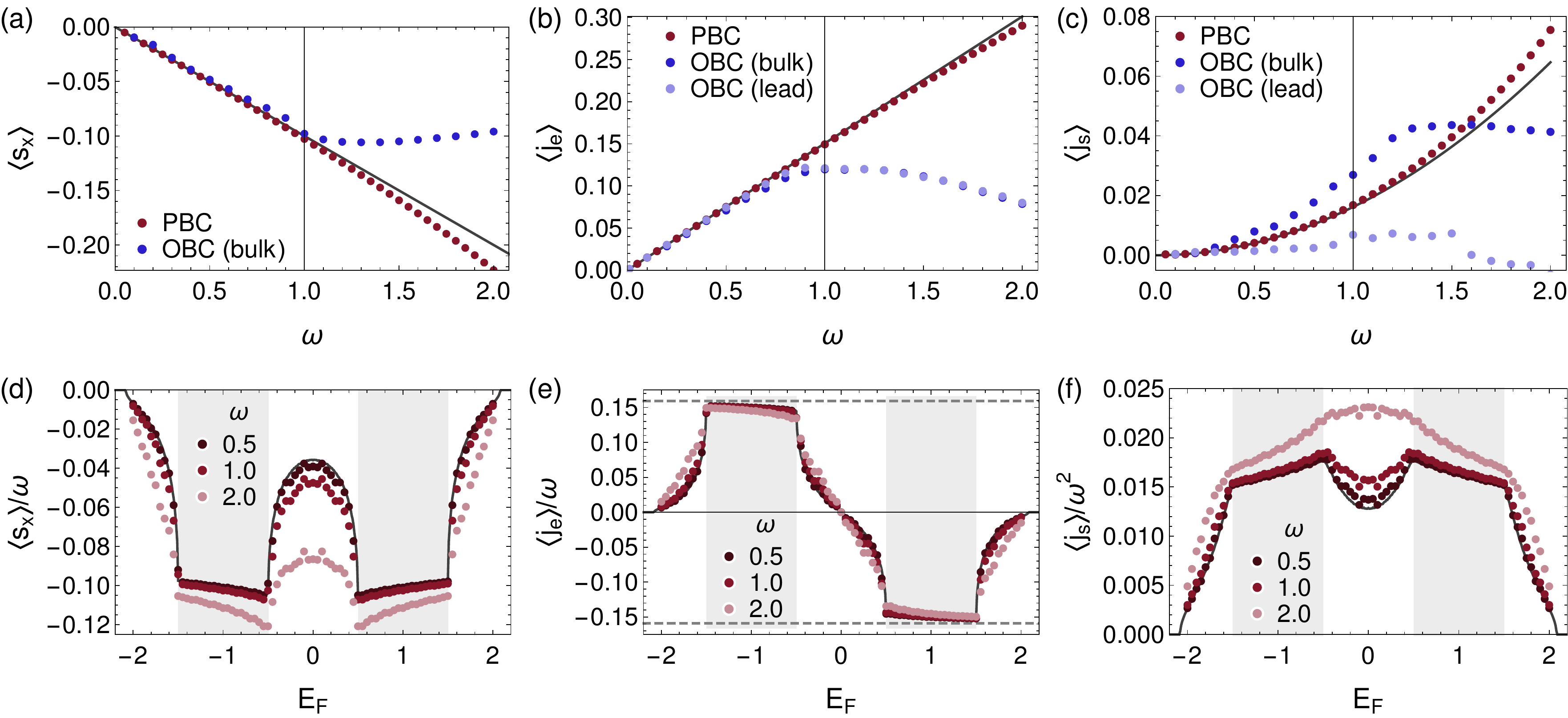}
    \caption{
    Averaged spin polarization $\langle s_x \rangle$, charge current $\langle j_e \rangle$, and spin current $\langle j_s \rangle$ as function of (a-c) the rotation frequency $\omega$ and (d-f) the Fermi energy $\EF$, respectively.
    In panels (a-c), the Fermi energy is fixed to $\EF/\tH=-1$.
    Dots are numerical data for periodic boundary conditions (PBC, red) and for finite size systems with non-magnetic leads attached (OBC, blue).
    For the latter, the observables are computed both in the bulk (darker blue) and the attached leads, far away from the magnetic system (lighter blue).
    Note that they mostly overlap in panel (b).
    The solid black line is the perturbative result, Eq.~\eqref{eq:expectationvalue}.
    The vertical line at $\omega=1$ indicates the driving frequency which corresponds to the SDW gap $\Delta_\text{\tiny{SDW}}=2J$.
    In panels (d-f), the driving frequency $\omega$ is fixed to $\omega=0.5, 1.0$, and $2.0$, see legends.
    Dots are numerical data for PBCs.
    The solid black line is the perturbative result, Eq.~\eqref{eq:expectationvalue}.
    The gray shaded areas indicate the SDW gap.
    Parameters for all panels are $\lambda/a=3$, $t_\text{\tiny{H}} = 1$, and $J/\tH = 0.5$.
    Details on the numerical calculations are provided in the main text.
    }
    \label{fig3}
}
\end{figure*}

In the following, we switch from the perturbative analysis in Sec.~\ref{sec:perturbationtheory} to non-perturbative numerics evaluation which also allows for treating other boundary conditions than periodic.
However, for simplicity, we fix the wavelength of the magnetic spiral to $\lambda/a = 3$ which does not qualitatively alter the results.

We numerically solve the time-dependent Schr\"odinger equation to simulate the dynamics of the electronic subsystem in real-space where we can easily evaluate the time- and space-dependent spin accumulation $s_x$, charge current $j_e$, and spin current $j_s$, see Eqs.~\eqref{eq:sx}-\eqref{eq:js}.
At the initial time $t=0$, the wavefunction of the electronic state $\psi_0=\psi(t=0)$ is obtained by diagonalizing the Hamiltonian $\mathcal{H}_0=\mathcal{H}(t=0)$, see Eq.~\ref{eq:H}, and filling all states up to the Fermi energy $\EF$.
We manually switch on the spiral motion $\omega\neq0$ at $t=0$ which evolves the system to a non-equilibrium state $\psi(t) = \hat{U}(t,0) \psi_0$.
Here, $\hat{U}(t', t)$ is a time-evolution operator from time $t$ to $t'$ and, for a short time period $\delta t$, it is expressed by
\begin{equation}
\hat{U}(t + \delta t, t) = \hat{T} \exp\left[-i \int_t^{t + \delta t} dt' \mathcal{H}(t')\right]
\label{eq:U} 
\end{equation}
where $\hat{T}$ is the time-ordering operator.
Employing the Suzuki-Trotter decomposition~\cite{Suzuki1994, Nakanishi1997, Misawa2019}, we numerically compute the time-evolution operator $\hat{U}$ and, finally, the nonequilibrium observables $\braket{\mathcal{O}(t)} = \langle\psi^\dagger(t)|\hat{O}|\psi(t)\rangle$.

Within the real-space scheme, we can apply different boundary conditions.
We use two different numerical codes for the simulations; (i) our self-written code and (ii) the open-source program T{\footnotesize KWANT}~\cite{Kloss2021}.
Our self-written code uses periodic boundary conditions (PBCs) or open boundary conditions (OBCs) but we can define different parameter regions, e.g., a spiral magnet region with polarized magnetic leads attached on both sides.
In this case, we minimize the effects of the boundary by choosing a sufficiently large system size.
For the case of simple non-magnetic leads attached to a spiral region, however, we use T{\footnotesize KWANT} which is more advanced.


\subsubsection{Leads attached and the limits of perturbation theory}
\label{sec:OBCnonmagnetic}

Using the non-perturbative numerical evaluations of the time-dependent Schr\"odinger equation, we can test the limits of our perturbation theory, Sec.~\ref{sec:perturbationtheory}.
Fig.~\ref{fig3} shows a comparison between the perturbative results (solid black lines), Eqs.~\eqref{eq:order1} and \eqref{eq:order2nd}, and non-perturbative numerical results for PBCs (red dots) or OBCs (blue dots).
Here, we refer to OBCs as attached half-infinite non-magnetic leads.
With OBCs, we evaluate the charge and spin currents $j_e$ and $j_s$ both in the bulk, i.e., as the average over the entire spiral region, and in the leads, i.e., as the average over \emph{both} attached metallic leads.
The definition of the latter is of significant importance as the spin currents on both ends, in general, have opposite sign, see Fig.~\ref{fig4} in the following section.

More precisely, Figs.~\ref{fig3} (a-c) show the DC component of the spin polarization $s_x$, charge current $j_e$, and spin current $j_s (= j_{s_x})$, respectively, as function of the driving frequency $\omega$.
For driving frequencies $\omega$ below the SDW gap, $|\omega| < 2|J|$, the perturbative results and numerical results are in quantitatively good agreement.
Also, the spin polarization $s_x$ and the charge current $j_e$, results for PBC and OBC do not differ significantly.
The average spin current $j_s$, on the other hand, depends on the type of boundary condition. 
It also depends strongly on the measurement position, i.e., inside the spiral region or outside in the attached lead, as the rotating spiral is a source of spin and continuously supplies angular momentum.
Note that a spin current is not conserved quantity in the spiral region due to broken spin rotational symmetry, while it is conserved in the leads.
Moreover, once the frequency $\omega$ is larger than the SDW gap, $|\omega| > 2|J|$, the results obtained for PBC and OBC start to deviate.
The results with PBCs are still close to the perturbative results, namely $s_x,j_e \propto \omega$ and $j_s\propto\omega^2$.
In turn, with OBCs, the pumped spin accumulation $s_x$ and charge and spin currents $j_e$ and $j_s$ saturate.

In Figs.~\ref{fig3}(d-f), we show the numerically computed susceptibilities related to the spin accumulation $s_x$, charge current $j_e$, and spin current $j_s$ as function of the Fermi energy $\EF$.
The driving frequency is fixed to $\omega=0.5, 1.0, 2.0$ and we only consider PBCs.
Again, we obtain good agreement between the perturbative and numerical results for low driving frequencies below the SDW gap, $|\omega| < 2|J|$.
For $|\omega| > 2|J|$ we enter the non-adiabatic regime where the perturbation theory breaks down and the non-perturbative numerical results start to deviate.
However, this deviation appears as a rather constant scaling factor, such that the main features of the pumped observables are conserved.


\subsubsection{Magnetic leads attached}
\label{sec:OBCmagnetic}
 
In the previous section, we have shown that the boundary condition can have a large impact on the pumped quantities since the rotating spiral is a source of angular momentum (spin). 
In this section, we investigate how the transport properties can be tuned if the attached leads are magnetic or non-magnetic.

Qualitatively, the impact of attached leads can be understood as follows.
The induced spin polarization $s_x$ inside the spiral region is fixed by the direction of rotation, i.e., the sign of $\omega$.
The rotating spiral generates spin, thus, at the interfaces $s_x$ may be transferred into the attached leads, leading to a spin current $j_s$ away from the spiral region.
So far, the phenomenology is similar to the spin pumping mechanism by precessing ferromagnets~\cite{Tserkovnyak2005a}.
However, simultaneously, the rotating spiral pumps charge $n_{e,i}=-e\braket{\vec{c}_i^\dagger\vec{c}_i}$.
With non-magnetic leads attached, the pumped spin-polarized electrons can easily be transferred through the leads.
In contrast, if the attached lead on the left/right side is ferromagnetic with magnetization $M^x_\mathrm{L/R}$, then states at the Fermi surface might be spin-polarized with $s^x_\mathrm{L/R} = M^x_\mathrm{L/R}$, depending on the Fermi level $\EF$.
In this case, only electrons with a spin-polarization $s^x_\mathrm{L/R}$ can be pumped into the lead or extract from it.
However, in order to transfer the spin accumulation $s_x$ from inside the spiral, electrons that are extracted from a lead must satisfy $s^x_\mathrm{L/R} || -s_x$, are then flipped to $s^x_\mathrm{L/R} || s_x$ by the rotating spiral, and can leave again on the other end only if $s^x_\mathrm{L/R} || M^x_\mathrm{L/R}$.
Otherwise, if one magnetic lead does not match these conditions it imposes a barrier for charge transport.
Then, the overall charge transport is suppressed as the total charge $e_\mathrm{tot}=-e\sum_i n_{e,i}$ is a conserved quantity.

For more quantitative results, we use our self-written code and simulate a composite structure which consists of two leads attached to either side of a magnetic spiral of length $L=30\lambda=90a$, which can each be magnetic or non-magnetic.
The lead regions are chosen sufficiently large that boundary effects on the time scale of the simulations can be neglected.
The magnetic exchange part of the Hamiltonian in Eq.~\eqref{eq:H} is then modified to
\begin{equation}
 \begin{split}
 \mathcal{H}_\mathrm{ex}(t) = 
 &- \sum_{i < -L/2}  J' \,\vec{c}_i^\dagger (\sigma_x M^x_\mathrm{L})\, \vec{c}_i \\
 &- \sum_{-L/2 \leq i < L/2}  J \,\vec{c}_i^\dagger (\bm{\sigma} \!\cdot\! \bm{M}(t))\, \vec{c}_i \\
 &- \sum_{i \geq L/2}  J' \,\vec{c}_i^\dagger (\sigma_x M^x_\mathrm{R})\, \vec{c}_i 
 \end{split}
\label{eq:HM} 
\end{equation}
where $\vec{M}(t)$ is the magnetic spiral given in Eq.~\eqref{eq:M} and depicted in Fig.~\ref{fig1}(b), $M^x_\mathrm{L/R}=-1, 0, 1$ is the magnetization in the left/right lead, and $J'$ is the exchange constant in the leads.
We achieve half-metallic spin-polarized leads by setting the exchange constant to $J'=2\tH$, assuring that the lower band is half-filled.

\begin{figure}
\centering{
    \includegraphics[width=\columnwidth]{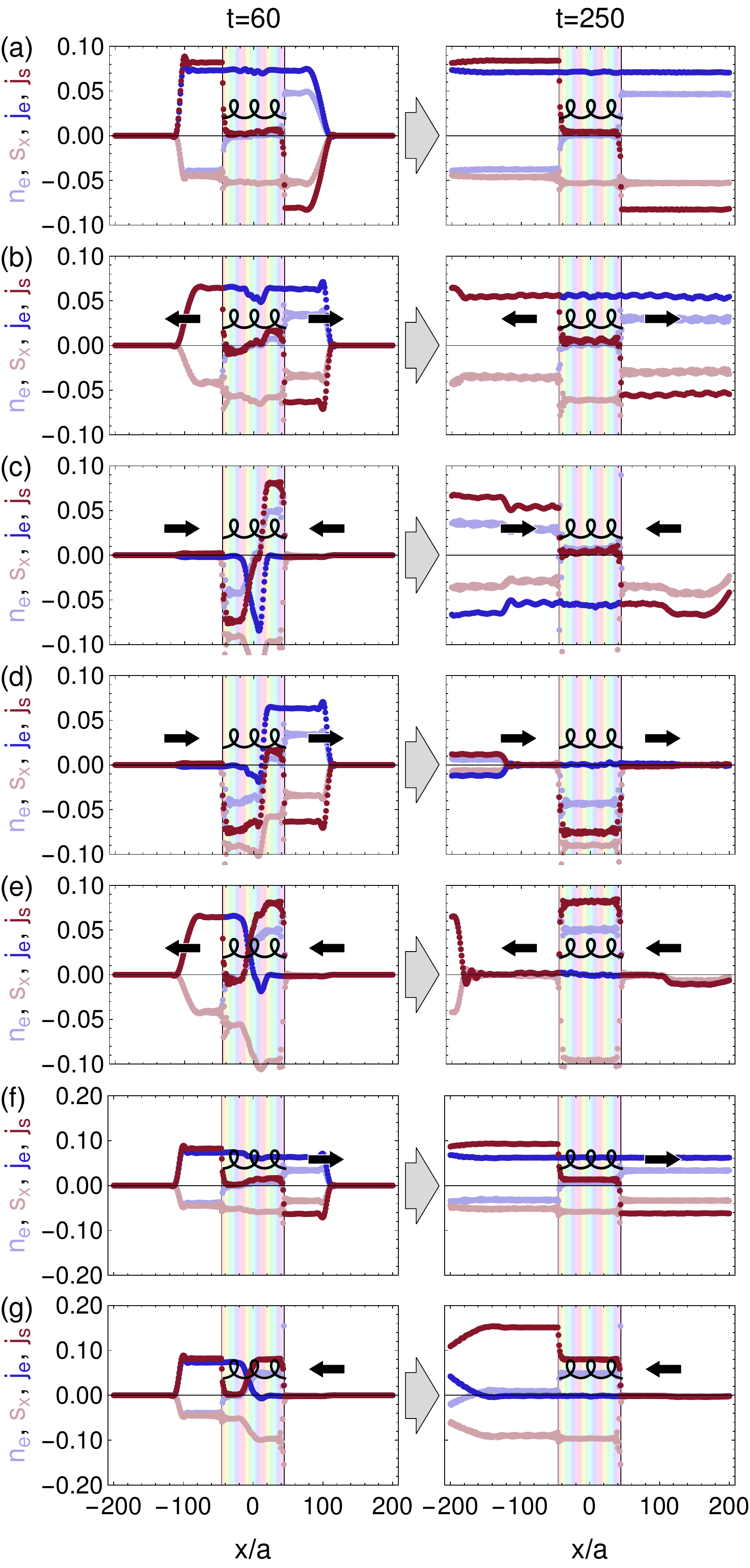}
    \caption{
    Real-space profiles of the charge density $n_e$ (light blue), spin accumulation $s_x$ (light red), charge current $j_e$ (blue) and spin current $j_s$ (red).
    Panels show numerical data at times $t=60$ (left column) and $t=250$ (right column) after spontaneously switching from $\omega=0$ to $\omega=0.5$ at $t=0$.
    As schematically indicated, the system consists of three regions:
    The central area of length $30\lambda=90a$ is the rotating helix from Fig.~\ref{fig1}.
    Leads attached on either side are (a) both non-magnetic, (b-e) both magnetic, or (f,g) mixed with one side non-magnetic and the other magnetic.
    The polarization of the magnetic leads is indicated by black arrows.
    For non-magnetic leads, the arrow is absent.
    Parameters for all panels are $\tH = 1$, $J = 0.5$, $J' = 2$, and $\EF=-1$.
    Offsets from $t<0$ have been subtracted.
    In some panels, data points for $n_e$ or $j_e$ are hidden behind the respective points for $s_x$ or $j_s$.
    For the full time-dependence, see also the Supplementary Movies 1-7.
    }
    \label{fig4}
}
\end{figure}

In Fig.~\ref{fig4}, we show the spatial profile of the charge and spin $n_e$ and $s_x$ (light blue and light red) as well as the charge and spin currents $j_e$ and $j_s$ (blue and red) obtained from our numerical simulations.
A constant offset corresponding to the profiles at $\omega=0$ is subtracted. 
We show the profiles for both a transient state at time $t=60$ (left column) and near the steady state ($t=250$, right column).
The spiral region is indicated by rainbows, similar to the spiral in Fig.~\ref{fig1}, and the magnetization $M_\mathrm{L/R}$ in the leads is indicated by solid black arrows or the absence of an arrow in the non-magnetic case ($M_\mathrm{L/R}=0$).
We discuss the resulting transient and steady states in the following.

\textit{Non-magnetic leads ---} 
To begin with, let us reconsider the case of non-magnetic leads on both sides, Fig.\ref{fig4}(a) or Supplementary Movie 1, which is in nice agreement with the qualitative discussion above. 
The spiral simultaneously pumps both charge $n_e$ and spin $s_x$.
The latter is equal on both sides of the spiral, while the charge $n_e$ is depleted on the left side and accumulated on the right side, reflected also in the spatially homogeneous current $j_e$.
This result is in agreement with the simple picture that the rotating right-handed spiral with $\omega>0$ is an Archimedean screw which transports spin-polarized electrons to the left.
The spin current $j_s$ reflects the fact that the spiral is a source of spin, leading to a dominant antisymmetric contribution to the spin current.
In addition, there is a small symmetric component.
Note that the latter is absent in magnetic leads, Fig.~\ref{fig4}(b-g), where spin transport by spin flips is suppressed and only transport by drifting spin-polarized electrons contributes, thus $n_e=\pm s_x$ and $j_e=\pm j_s$.

\textit{Magnetic leads (antiparallel) ---} 
Next, let us consider half-filled spin-polarized magnetic leads with opposite polarizations, $M_\mathrm{L}=-M_\mathrm{R}$, attached to both ends of the spiral.
Fig.~\ref{fig4}(b) or Supplementary Movie 2 shows the result for lead polarizations which both match the spiral pumping of the case under consideration, i.e., $M_\mathrm{L}=-1$ and $M_\mathrm{R}=1$ for $\eta=1$, $\omega>0$, and $\EF<0$.
Electrons with $s_x>0$ are pumped from the right lead into the spiral, resulting in $n_e>0$ and $s_x<0$. 
At the interface to the spiral region, their spin is flipped to $s_x<0$ such that, on the left end, they can be pumped into the lead with $M_\mathrm{L}<0$.
In Fig.~\ref{fig4}(c) or Supplementary Movie 3, the lead polarizations have been switched such that both leads are now blocking.
Accordingly, neither charge nor spin can initially leave the spiral region on either end, see panel for $t=60$.
However, as electrons/holes accumulate on the left/right side within the spiral region, they eventually fill enough states to activate charge transport in the inverse direction, c.f. Fig.~\ref{fig2}(b).
Once the direction of charge transport is reversed, the blockade breaks down as the system corresponds to the space-inversion symmetric partner of Fig.~\ref{fig4}(b).

\textit{Magnetic leads (parallel) ---} 
If both half-filled spin-polarized magnetic leads have the same polarization, $M_\mathrm{L}=M_\mathrm{R}$, see Fig.~\ref{fig4}(d,e)  or Supplementary Movies 4 and 5, then only one end of the spiral region has matching conditions.
On short time scales, a shock wave of charge $n_e>0$ and spin $s_x<0$ can leave the spiral region on the right side for $M_\mathrm{R/L}=1$ (left side with $n_e<0$ for $M_\mathrm{R/L}=-1$).
However, transport is not possible in the steady state as the other end of the spiral is blocking charge transfer, leaving the spiral region discharged (or charged) in the steady state, without any continued transport.

\textit{One magnetic and one non-magnetic lead ---} 
In case that one attached lead is non-magnetic, here $M_\mathrm{L}=0$, only the other lead imposes boundary conditions on the transport.
Fig.~\ref{fig4}(f) or Supplementary Movie 6 show the result if the lead on the right side is magnetic and matches the transport conditions, $M_\mathrm{R}=1$.
Just as for matching lead conditions, c.f. Fig.~\ref{fig4}(b), as steady charge and spin current persists.
On the magnetic side, again, the transport is due to the drift motion of spin-polarized electrons, thus $n_e=-s_x$ and $j_e=-j_s$.
In contrast, on the non-magnetic side, similar to the completely non-magnetic case in Fig.~\ref{fig4}(a), the spin polarization is larger due to spin flips.
Finally, in Fig.~\ref{fig4}(g) or Supplementary Movie 7 the magnetic lead does not match the transport conditions of the rotating spiral, $M_\mathrm{R}=-1$, i.e., is blocking charge transport.
Similar to Fig.~\ref{fig4}(e) with one matching and one blocking lead polarization, an initial shock wave of electrons is emitted from the spiral region into the matching condition (non-magnetic) lead.
As the other end of the spiral is blocking charge transport, however, this initial process soon ends and in the steady state, the charge current stops, $j_e=0$.
For the spin current, the situation is different as spin is not conserved.
Here, on the one side, the spin transport into the magnetic lead is blocked as both, pumping polarized electrons and propagating spin flips, are gapped out.
On the other side, conduction electrons are not available on in the steady state, as explained before, but spin transport by spin flips is still an option.
Thus, in our simulations, we observe a remarkably high spin current (and polarization) into the non-magnetic lead.

In summary, attaching magnetic and non-magnetic leads can drastically alter the spin and charge transport properties of the spinning spiral magnet and, besides symmetric transport of both quantities, realize charge and spin diodes and switches or spin sources.


\subsection{Effects of disorder}
\label{sec:disorder}
 
\begin{figure*}
\centering{
    \includegraphics[width=\textwidth]{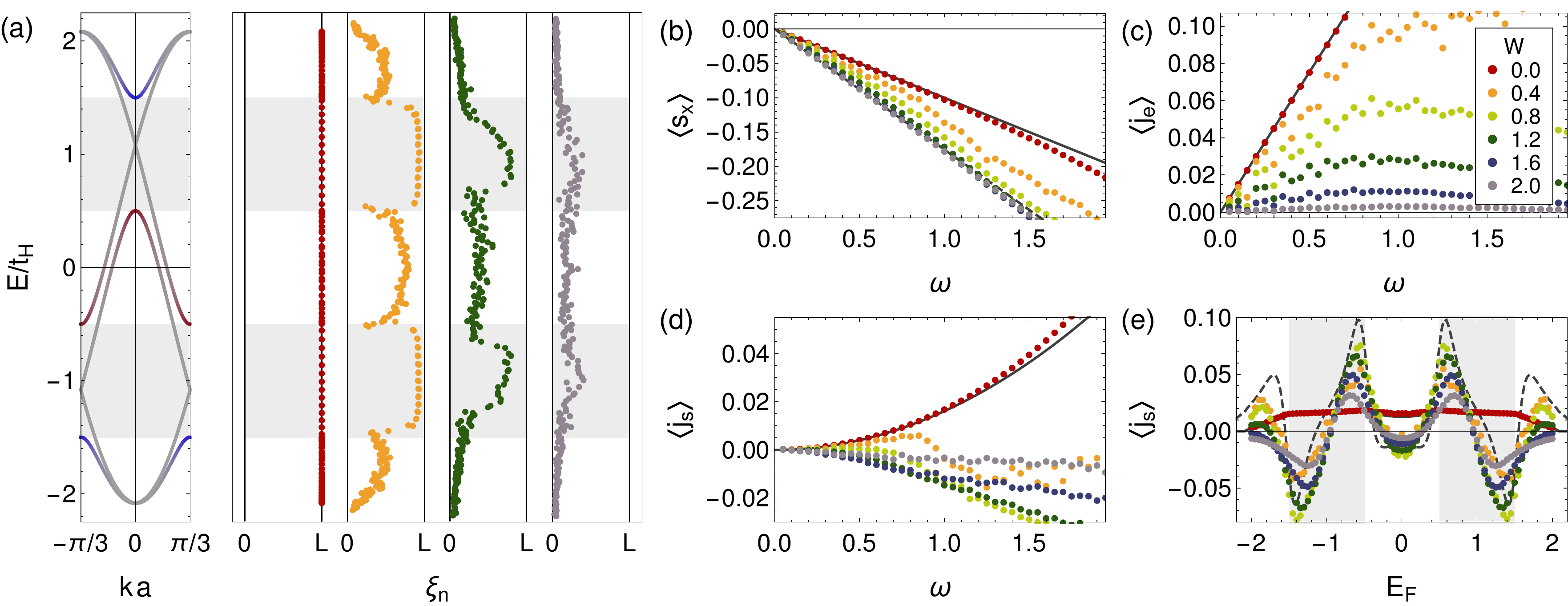}
    \caption{
    (a) Band structure for $\lambda/a = 3$, c.f. Fig.~\ref{fig1}, and localization length $\xi_n$ for disorder strengths $W=0, 0.4, 2.0$.
    Panels (b-e) show the effect of disorder $W=0, 0.4, 0.8, 1.2, 1.6, 2.0$ on (b) the spin polarization, (c) the charge current, and (d-e) the spin current as function of (d) the driving frequency $\omega$ or (e) the Fermi energy $\EF$.
    	For the color of data points, see the inset of panel (c).
	Solid lines are results from our perturbation theory, see Eqs.~\eqref{eq:order1} and \eqref{eq:order2nd} and c.f. Fig.~\ref{fig3}(a-c) and (f).
	The dashed line in (e) is the contribution of the dissipative susceptibility, Eq.~\eqref{eq:order2d}.
	For all panels, the system size is $L=50\lambda=150a$ with periodic boundary conditions.
	Parameters are $t_\text{\tiny{H}} = 1$, $J = 0.5$, $\EF = -1$, $\omega=1$, and $\tau = 2.5$, if not stated otherwise.
    }
    \label{fig5}
}
\end{figure*}

While the above results and discussions considered clean samples and ideal model systems, disorder is expected to tremendously impact the pumping phenomena.
In the following, we present our numerically obtained results for the spin accumulation $s_x$, charge current $j_e$, and spin current $j_s$ in the presence of non-magnetic impurities and link these results to the localization length $\xi$ and second order dissipative susceptibility $\chiD$ of the spin current.

To study the effects of disorder represented by non-magnetic impurities, we consider a model system of size $L=50\lambda=150a$ with periodic boundary conditions.
The Hamiltonian $\mathcal{H}'$ of this dirty model system consists of two parts,
\begin{equation}
 \mathcal{H}' = \mathcal{H} + \mathcal{W} = \mathcal{H} + \sum_i W_i \vec{c}_i^\dagger \vec{c}_i \,\,,
 \label{eq:Hdirty}
\end{equation}
where $\mathcal{H}$ is the Hamiltonian of the clean system, see Eq.~\eqref{eq:H}, and $\mathcal{W}$ is a non-magnetic random potential.
We define $\mathcal{W}$ by the uniformly distributed random onsite energy $W_i \in \left[-W/2, W/2\right]$ with the spatial averages $\braket{W_i} = 0$ and $\braket{W_i W_j} = \delta_{i\!j} W^2/12$.
For all following results, we average over $200$ different disorder configurations.

First, we briefly analyze the impact of non-magnetic disorder on the static properties of the electronic system.
We define the localization length $\xi_n$ of the $n$-th eigenstate $\psi'_n$ of $\mathcal{H}'$ via the inverse of the inverse participation ratio $I_n$, i.e.,
\begin{equation}
 \xi_n = I_n^{-1} = a \left( \sum_i |\psi_{n,i}|^4 \right)^{-1} \,\,,
 \label{eq:xi}
\end{equation}
where the sum is over all lattice sites $i$. 
By numerically diagonalizing the Hamiltonian $\mathcal{H}'$, the localization length $\xi_n$ is evaluated for several different disorder strengths up to $W=2$, see Fig.~\ref{fig5}(a). 
With disorder absent, $W=0$, all states are fully extended over the system.
Once disorder is introduced, $W>0$, the localization length $\xi_n$ monotonically decreases as the disorder strength $W$ increases.
The helical channels within the SDW gap are more robust against disorder due to the spin-momentum locking, however, following the discussion in Sec.~\ref{sec:perturbationtheory}, they do not contribute to the charge current $j_e$.

Next, we numerically evaluate the spin accumulation $s_x$, charge current $j_e$, and spin current $j_s$, using our self-written code from Sec.~\ref{sec:numerics} with disordered Hamiltonian $\mathcal{H}'$, Eq.~\eqref{eq:Hdirty}, and by taking again the disorder average over $200$ disorder configurations and time average after an initial relaxation time.
The response of the spin accumulation $s_x$ and charge current $j_e$ to disorder are just opposite: 
The absolute value of the spin accumulation $s_x$ increases with increasing disorder $W$ until it saturates, see Fig.~\ref{fig5}(b), captured by renormalization of the first order susceptibility $\chi^{(1)}_{j_e}$. 
In a strongly disordered regime, the response is described by single-site (2-level) problem as most of states are localized and given by $\langle s_x \rangle = -C \omega/(4J)$ where $C \approx 0.351$ is a statistical factor of 1/3-filling and $L = 150a$.
The charge current $j_e$ decays monotonically with increasing disorder $W$, see Fig.~\ref{fig5}(c), due to the increased localization of states, $\xi_n \to 0$.
Notably, in contrast to these monotonic behaviors, the spin current $j_s$ can behave non-monotonically as function of both driving frequency $\omega$, Fig.~\ref{fig5}(d), and Fermi energy $\EF$, Fig.~\ref{fig5}(e), including possible sign changes as function of both these control parameters.
For small disorder $W\ll1$ and driving frequency $\omega\ll1$ the spin current starts off positively, $j_s>0$, as described by our second order non-dissipative perturbation theory, Eq.~\eqref{eq:order2nd}, see solid line in Fig.~\ref{fig5}(e).
However, at stronger disorder $W$ and frequency $\omega$, additional contributions become relevant and soon also dominant.
These additional contributions are strongest at the edges of the SDW gaps and induce a sign change across the SDW gap, in agreement with our second order dissipative perturbation theory, Eq.~\eqref{eq:order2d}, see dashed  line in Fig.~\ref{fig5}(e).
Note also that the dissipative contribution, which stems from the long-ranged helical channels, survives even in the strong disorder case $W=2$ where the charge current $j_e$ is already suppressed.
Vice versa, the dissipative contribution is suppressed in the clean limit $W\to0$ where the scattering lifetime diverges, $\tau\to\infty$.

In summary, adding non-magnetic disorder to our model may enhance spin-related pumping phenomena whereas the pumped charge current is suppressed.
However, here, we did not consider magnetic impurities or spatial/temporal defects in the  phason $\phi(x,t)$ and our analysis was restricted to a purely one-dimensional model.


\section{Summary and Discussion}
\label{sec:discussion}

We have investigated the charge and spin pumping phenomena in a rotating/translating one-dimensional spiral magnet \emph{without} spin-orbit coupling (SOC).
We studied all regimes of spiral periods $\lambda$, in particular focussing on the case of atomically short pitches, $\lambda\sim a$, but discussed also the two limiting cases, namely the collinear antiferromagnet, $\lambda=2a$, and the long wavelength (ferromagnetic) limit, $\lambda\gg a$.
Our analytical and numerical results predict in detail the non-trivial charge and spin transport properties of these systems.

The rotation of a magnetic spiral at angular frequency $\omega$ -- equivalent to a translation at velocity $v$ -- was predicted to pump an electrical current $\vec{j}_e$ in the direction of the $\vec{Q}$-vector, recently interpreted by del Ser \textit{et al.}~\cite{delSer2021} as an electromagnetic Archimedean screw.
This picture agrees with simple symmetry arguments which predict $|\vec{j}_e| \propto \omega$.
However, we find a non-trivial dependence on the spiral wavelength $\lambda$, showing a maximum at around $\lambda \sim 3.5a$ and vanishing in both the antiferromagnetic and ferromagnetic limit.
This behavior is explained by the Berry curvature $\mathcal{B}_{k,\phi}$ in mixed momentum-phason space which determines the current $\vec{j}_e$.
Concentration of the Berry curvature in the SDW channels leads to constant plateaus inside the SDW gaps, since the gapless helical channels do not mediate the charge transport.
Nevertheless, the gapless channels provide a route for the cancellation of Berry curvature once $\lambda$ increases: In the limit of large wavelengths $\lambda\gg a$ we obtain $j_e \propto \omega/\lambda \propto - v/\lambda^2$, spoiling the oversimplified picture of a classical Archimedean screw~\cite{delSer2021,Rorres2000} or topological Thouless pumping~\cite{Thouless1983,Niu1984}.
However, to this point, we did not yet include SOC in our theory which could open additional gaps and possibly prevent the cancellation of Berry curvature.

In addition to the electrical current $\vec{j}_e$, the rotation generates a spin accumulation $\vec{s}$ and spin current $\vec{j}_{\vec{s}}$.
The spin $\vec{s}$ points perpendicular to the rotation plane of the spin spiral, which is determined by the magnetic properties of the system.
The pumping direction of the spin current $\vec{j}_{\vec{s}}$, however, is again set by the direction of the $\vec{Q}$-vector of the magnetic spiral.
In contrast to the electric current $\vec{j}_e$, the spin accumulation and current are even function of the Fermi energy $\EF$.
The spin accumulation $\vec{s}\propto\omega$ is approximately constant over all ranges of $\lambda$, whereas $\vec{j}_{\vec{s}}\propto\omega^2$ is composed of a (non-dissipative) Fermi sea component and a (dissipative) Fermi surface component with distinct properties.
However, similar to $\vec{j}_e$, also $\vec{j}_{\vec{s}}$ vanishes in both the antiferromagnetic and ferromagnetic limit but peaks around $\lambda \sim10a$.

We have also studied the effects of attached non-magnetic or spin-polarized half-metallic magnetic leads as well as the impact of non-magnetic disorder.
We find that the locking between the charge current $\vec{j}_e$ and spin current $\vec{j}_{\vec{s}}$ inside the rotating spiral imposes boundary conditions on the global transport if leads are attached.
As a result, charge and spin transport might be simultaneously enabled.
In addition, we can realize scenarios where only the charge current is blocked or inverted, or where both currents are blocked.
Control over the properties of attached leads therefore could be used to construct charge and spin diodes and rectifiers.
However, we also show that the imperfectly quantized charge pumping is suppressed by the non-magnetic disorder, in contrast to the spin accumulation which is enhanced.
The effects of disorder on the spin current are less trivial as the finite excitation lifetimes mix the dissipative and non-dissipative contributions, leading enhanced spin currents and eventually sign changes at the edges of the SDW gaps.
Noteworthy, this modified spin current of mostly dissipative origin persists at much larger disorder strengths than the electrical current.
Therefore, future investigations should also consider effects of magnetic disorder and fluctuations in the spin spiral, as well as the crossover between our theory and the SOC-driven transport in the dirty limit which was discussed by del Ser \textit{et al.}~\cite{delSer2021}.
Also all thermal effects have been neglected in our study which might be the topic of future studies.

Finally, let us comment on possible experimental realizations.
The pumped charge and spin currents change their sign depending on the helicity $\eta$ of the spiral, suggesting that the transport phenomena are suppressed for multichiral systems.
Chiral magnets come with a well-defined helicity which is fixed by the underlying crystal structure.
But, as the spiraling Dzyaloshinskii-Moriya interaction requires spin-orbit coupling, it is generally small and the wavelengths with $\lambda\gtrsim20\rm{nm}$ rather large~\cite{Nakanishi1980, Bak1980}.
Moreover, the spiral phase naturally comes in domains of different orientations~\cite{Schoenherr2018} which first need to be properly combed~\cite{Masell2020}.
Alternatively, magnets with frustrated interactions, e.g. RKKY interaction, can be considered as they offer atomically small wavelengths, $\lambda\lesssim5\rm{nm}$.
Such systems are multichiral but they can be poled using the electrical magnetochiral effect~\cite{Jiang2020}.
For rotating/driving the spin spiral, the authors of Ref.~\cite{delSer2021} suggested to use an oscillating external magnetic field perpendicular to the $\vec{Q}$-vector which is, however, only a 2nd order effect in the driving field.
We suggest to also reconsider that a magnetic field is already the generator of a spin rotation, thus an oscillating magnetic field perpendicular to the spiral plane can be used to detect the AC equivalent of our theory.
The rotation induced by a constant magnetic field would be damped out after some time.
However, we can counteract by replacing the magnetic field with a continuous source of spin, e.g., by exploiting the spin Hall effect in an adjacent heavy metal layer for spin-charge conversion, as frequently exploited in spin orbit-torque setups in thin films.
In either case, we predict that the resulting pumped charge current density $\braket{j_{e, 3D}}(\omega) = \braket{j_{1D}}/a^2 = e \omega/(2\pi l_a)$ can be of the order of $j_e=2 \times 10^5\rm \ A/m^2$, assuming the spiral magnet is the frustrated Kagom\'e magnet $\rm Gd_3Ru_4Al_{12}$ which has $\lambda \sim 2.8$ nm and $a \sim 5 \AA$~\cite{Hirschberger2019}, and the spiral moves at a reasonably slow velocity of $v = 1\ \rm cm/s$, i.e., $\omega = 2.3$ MHz.


\acknowledgements{
D.K. and Y.L. were supported by the RIKEN Special Postdoctoral Researcher Program (SPDR).
J.M. acknowledges financial support by JSPS (project No. 19F19815) and the Alexander von Humboldt foundation.
N.N. was supported by JST CREST Grant Number JPMJCR1874 and JPMJCR16F1, Japan, and JSPS KAKENHI Grant number 18H03676.
}

\appendix
\section{Details on perturbation theory}
\begin{figure}[tbp]
\centering
\includegraphics[width = \columnwidth]{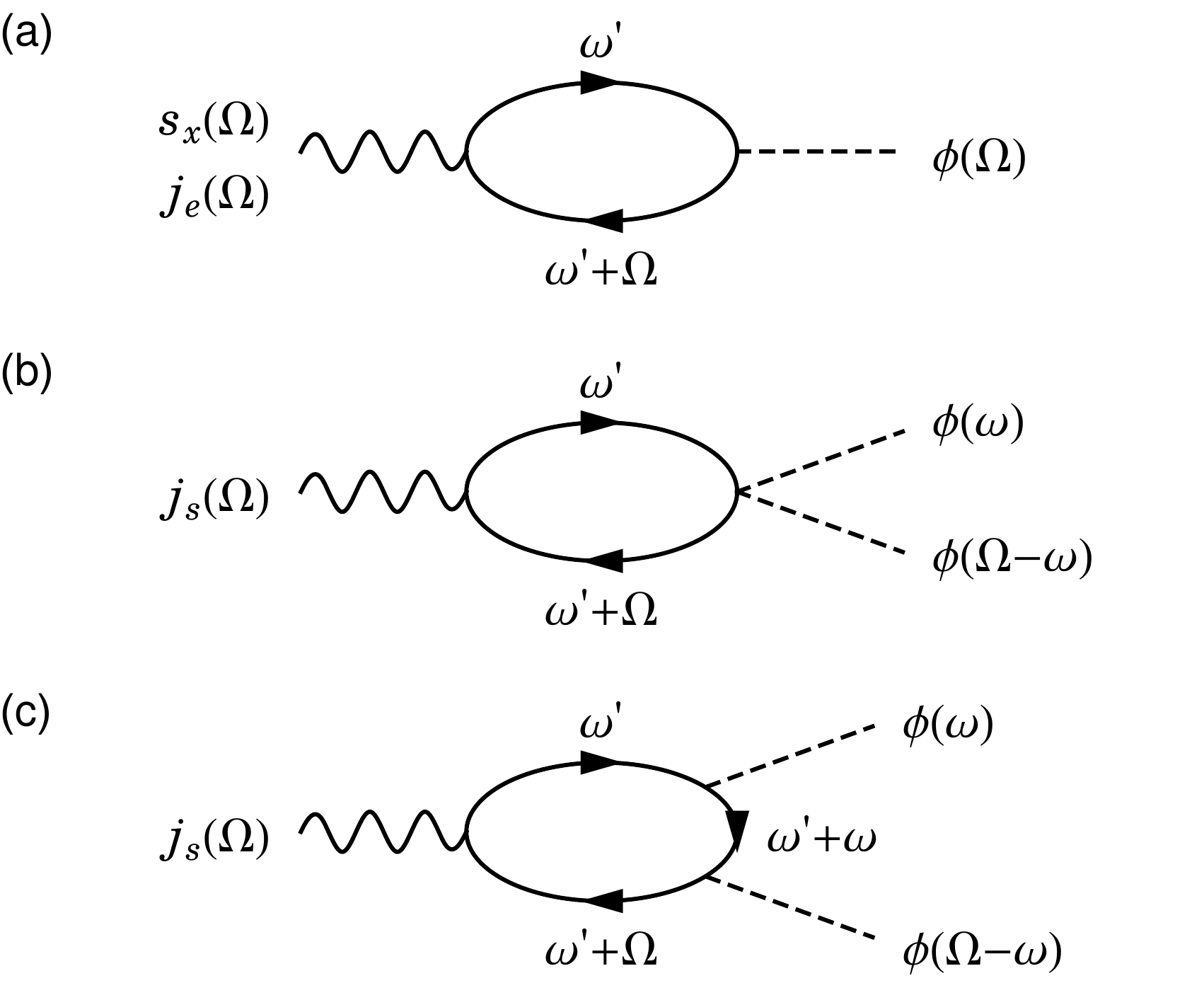}
\caption{
The Feynman diagrams describing the (a) first- and (b, c) second-order responses.
The wiggled line is the observable, the circle is the particle-hole bubble with the solid line being the electron propagator, and the dashed lines are phason couplings.
}
\label{fig_FD}
\end{figure}

\label{sec:perturbationdetails}
The nonequilibrium physical observable is evaluated as
\bea
\braket{O} (t) &=& \left. -i \int \frac{dq}{2\pi} {\rm Tr}\left[\hat{O} \hat{G}_q^<(t,t')\right] \right|_{t' \rightarrow t},
\label{eq_a1}
\eea
where $\hat{G}_q^<(t,t') = i \braket{\bm c^\dagger_q(t') \bm c_q(t)}$ is the lesser Green's function on the Keldysh contour~\cite{Haug2008}.
By expanding the contour-ordered Green's function up to the second-order with respect to the phason variable,
one can obtain that
\bea
\nonumber
&&\hat{G}_q(\tau ,\tau') = \hat{g}_q (\tau - \tau')\\
\nonumber
&& - J\int_C d\tau_1 \hat{g}_q (\tau - \tau_1) \phi(\tau_1) \hat{\Sigma}_\phi^{(1)} \hat{g}_q (\tau_1 - \tau')\\
\nonumber
&& -\frac{J}{2} \int_C d\tau_1 \hat{g}_q (\tau - \tau_1) \phi(\tau_1)^2 \hat{\Sigma}_\phi^{(2)} \hat{g}_q (\tau_1 - \tau') \\
\nonumber
&& + J^2  \int_C d\tau_1 \int_C d\tau_2 \hat{g}_q (\tau - \tau_1) \phi(\tau_1) \hat{\Sigma}_\phi^{(1)} \\
&&\hspace{2cm} \times \hat{g}_q (\tau_1 - \tau_2)\phi(\tau_2) \hat{\Sigma}_\phi^{(1)} \hat{g}_q (\tau_2 - \tau'),
\label{eq_a2}
\eea
where $\hat{g}_q (\tau)$ is the Green's function of the bare Hamiltonian, $\mathcal{H}_0$.
The nonequilibrium physical observables are obtained by substituting Eq.~(\ref{eq_a2}) into Eq.~(\ref{eq_a1}) as
\bea
\nonumber
&&\braket{O}(t) = \\
\nonumber
&& \left. iJ \int \frac{dq}{2\pi} \int dt_1 {\rm Tr}\left[
\hat{O} \hat{g}_q (t - t_1) \hat{\Sigma}_\phi^{(1)} \hat{g}_q (t_1 - t')
\right]^< \right|_{t' \rightarrow t}\phi(t_1)\\
\nonumber
&& +\frac{iJ}{2} \int \frac{dq}{2\pi} \int dt_1 \left.  {\rm Tr}\left[
\hat{O} \hat{g}_q (t - t_1) \hat{\Sigma}_\phi^{(2)} \hat{g}_q (t_1 - t')
\right]^< \right|_{t' \rightarrow t}\phi(t_1)^2  \\
\nonumber
&& -iJ^2 \int \frac{dq}{2\pi} \int dt_1 \int dt_2 {\rm Tr}\left[
\hat{O} \hat{g}_q (t - t_1) \hat{\Sigma}_\phi^{(1)} 
\right.\\
&&\hspace{1.5 cm} \left.\left.  \times
\hat{g}_q (t_1 - t_2) \hat{\Sigma}_\phi^{(1)} \hat{g}_q (t_2 - t')
\right]^< \right|_{t' \rightarrow t}\phi(t_1) \phi(t_2),
\label{eq:green_function}
\eea
where we have dropped a term independent of the phason variable.
The first term of Eq.~\eqref{eq:green_function} yields the linear-order response to the phason variable contributing to the spin polarization and the charge current, whereas the second and third terms give the second-order response responsible for the spin current.

\subsection{Linear-order Response}
The first-order response to the phason degrees of freedom described by the Feynman diagram in FIG.~\ref{fig_FD} (a) contributes to the spin polarization and the charge current.
In the linear-order response is evaluated as
$
\braket{O}(t) = \int \frac{d\omega}{2\pi} e^{-i\Omega t} K_{O}^{(1)} (\Omega) \phi(\Omega),
$
where $K_{O}^{(1)}(\Omega)$ is the first-order response function.
With the analytical continuation performed as $\int dt_1 [A(t, t_1) B(t_1, t')]^< = \int dt_1 \left[A^r(t, t_1) B^<(t_1, t')\right.$ $\left.+ A^<(t, t_1) B^r(t_1, t')\right]$~\cite{Haug2008}, the response function is evaluated as
\bea
\nonumber
K_O^{(1)} (\Omega) &=& iJ \int \frac{dq}{2\pi} \int \frac{d\omega'}{2\pi} \left(\hat{O}\right)_{ij} \left(\hat{\Sigma}_\phi^{(1)}\right)_{ji}\\
\nonumber
&&\times
\left[
g^r_{q,j}(\omega') g^<_{q,i}(\omega' + \Omega) + g^<_{q, j}(\omega') g^a_{q,i}(\omega' + \Omega)
\right]\\
\nonumber
&\approx&- \Omega J \int \frac{dq}{2\pi} \frac{\left(\hat{O}\right)_{ij} \left(\hat{\Sigma}_\phi^{(1)}\right)_{ji}}{\left(\epsilon_{q, i} - \epsilon_{q, j}\right)^2}
\left[f(\epsilon_{q, i} - \EF) - f(\epsilon_{q, j} - \EF)\right]\\
\nonumber
&\equiv&  -i\Omega \chi_{O}^{(1)}\\
\chi_{O}^{(1)} &=& 2J\int \frac{dq}{2\pi} \sum_{i \ne j}  \frac{f(\epsilon_{q, i} - \EF)}{\left(\epsilon_{q, i} - \epsilon_{q, j}\right)^2} {\rm Im}\left[ \left(\hat{O}\right)_{ij} \left(\hat{\Sigma}_\phi^{(1)}\right)_{ji}\right]
\eea
where $\psi_0^{q, i}$ and $\epsilon_{q, i}$ denote an eigenstate and eigenvalue of the unperturbed Hamiltonian $\mathcal{H}_0$, 
$i$ is a band index,
$\left(\hat{O}\right)_{ij} = \braket{\psi_{0}^{q, i} | \hat{O} |\psi_{0}^{q, j}}$ is a matrix element of the operator $\hat{O}$,
$g_{q, i}^{r (a)}(\omega) = \left( \omega + \EF - \epsilon_{q, i} \pm \frac{i}{2\tau}\right)^{-1}$ is the bare retarded (advanced) Green's function, 
$g_{q, i}^{<}(\omega) = if(\omega) A_{q, i}(\omega)$ is the bare lesser Green's function,  and
$A_{q, i} = i\left[ g_{q, i}^{r}(\omega) - g_{q, i}^{a}(\omega) \right]$ is the spectral function.
Note that the zeroth order terms in the frequency $\omega$ vanishes due to the gauge invariance.

\subsection{Second-order response}
The spin current is yielded as the second-order response to the phason variable.
In the second-order, there are two distinct contributions arising from the second and third term of Eq.~\eqref{eq:green_function}.
In analogy to electronic conductivity, we refer them the diamagnetic and paramagnetic contributions, respectively.
Each contribution of the spin current is represented as FIG.~\ref{fig_FD} (b, c) and evaluated by
\bea
\nonumber
\braket{j_s}(\Omega) = \int d\omega \left[ K_d(\Omega) + K_p(\Omega, \omega) \right] \phi(\omega) \phi(\Omega - \omega),\\
\label{eq:appendix_js}
\eea
where $\braket{j_s}(\Omega)$ is the Fourier component of the spin current defined by $\braket{j_s}(t) = \int \frac{d\Omega}{2\pi} e^{-i\Omega t} \braket{j_s}(\Omega)$, and $K_{d (p)}$ is the diamagnetic (paramagnetic) response function defined by
\bea
\nonumber
K_d(\Omega) &=& \frac{iJ}{2} \int \frac{dq}{2\pi} \int \frac{d\omega'}{2\pi} 
{\rm Tr}\left[\hat{j}_s \hat{g}_q (\omega') \hat{\Sigma}_\phi^{(2)} \hat{g}_q(\omega' + \Omega)  \right]^<,\\
\label{eq:dmag}\\
\nonumber
K_p(\Omega, \omega) &=& -iJ^2 \int \frac{dq}{2\pi} \int \frac{d\omega'}{2\pi} {\rm Tr}\left[
\hat{j}_s \hat{g}_q (\omega') \hat{\Sigma}_\phi^{(1)} \right.\\
&&\left. \hspace{1.5cm}
\times \hat{g}_q (\omega' + \omega') \hat{\Sigma}_\phi^{(1)} \hat{g}_q (\omega' + \Omega) 
\right]^<.
\label{eq:pmag}
\eea

\subsubsection{Cancellation of gauge-dependent terms}
Since our focus on this paper is the DC transport ($\Omega \rightarrow 0$), the response functions can be expanded with respect to the $\omega$ and $\Omega$, which yields
\bea
K_d(\Omega) &\approx& \chi_d^{(0)} -i\Omega \chi_d^{(1)} - \frac{\Omega^2}{2} \chi_d^{(2)},
\label{eq:grad1}
\\
\nonumber
K_p(\Omega, \omega) &\approx& \chi_p^{(0, 0)} -i\Omega \chi_p^{(1, 0)} -i\omega \chi_p^{(0, 1)}\\
&&
\hspace{-0.3cm}
- \frac{1}{2}\left[\Omega^2 \chi_p^{(2, 0)} + 2 \Omega \omega \chi_p^{(1,1)} +  \omega^2 \chi_p^{(0, 2)}\right],
\label{eq:grad2}
\eea 
where the expansion coefficients are defined as $\chi_d^{(n)} \equiv \partial_{\Omega}^n K_d(\Omega)|_{\Omega \rightarrow 0}$ and $\chi_p^{(n, m)} \equiv \partial_{\Omega}^n \partial_\omega^{m} K_p(\Omega, \omega)|_{\Omega, \omega \rightarrow 0}$. 
By substituting Eq.~(\ref{eq:grad1},~\ref{eq:grad2}) into Eq.~\eqref{eq:appendix_js}, one can obtain that
\bea
\nonumber
\braket{j_s}(t) &=& \left(\chi_d^{(0)} + \chi_p^{(0, 0)}\right) \phi^2(t)\\
\nonumber
&&+ \left(2 \chi_d^{(1)} + 2 \chi_p^{(1, 0)} + \chi_p^{(0, 1)}\right) \phi(t)\frac{d\phi(t)}{dt}\\
\nonumber
&&+ \frac{1}{2}\left(2 \chi_d^{(2)} + 2 \chi_p^{(2, 0)} + \chi_p^{(1, 1)} + \chi_p^{(0, 2)}\right) \phi(t)\frac{d^2\phi(t)}{dt^2}\\
&&+ \frac{1}{2}\left(2 \chi_d^{(2)} + 2 \chi_p^{(2, 0)} + \chi_p^{(1, 1)}\right) \left(\frac{d\phi(t)}{dt}\right)^2,
\label{eq:append_js}
\eea
where the first three terms explicitly depend on the phason variables, $\phi(t)$, therefore, forbidden by the U(1) symmetry.
In the following, we will show that the gauge-dependent terms exactly vanish.

Recalling that $\hat{\Sigma}_\phi^{(1)} =-\frac{1}{J} \left( \frac{\partial \hat{H}}{\partial \phi_0} \right)$ and $\hat{\Sigma}_\phi^{(2)} = -\frac{1}{2J}\left( \frac{\partial^2 \hat{H}}{\partial \phi_0^2} \right)$, 
one can show the relationship between the diamagnetic and paramagnetic contributions as
\bea
\nonumber
&&K_d(\Omega) = -\frac{i}{4} \int\frac{dq}{2\pi} \int\frac{d\omega'}{2\pi} {\rm Tr}\left[
\hat{j}_s \hat{g}_q (\omega') \left( \frac{\partial^2 \hat{H}}{\partial \phi_0^2} \right) \hat{g}_q(\omega' + \Omega)  
\right]^<\\
\nonumber
=&&- \frac{iJ}{4} \int\frac{dq}{2\pi} \int\frac{d\omega'}{2\pi} {\rm Tr}\left[
\hat{j}_s \partial_\phi \hat{g}_q (\omega') \hat{\Sigma}_\phi^{(1)}  \hat{g}_q(\omega' + \Omega)
\right.\\
\nonumber
&&\left. \hspace{3cm}
+ \hat{j}_s \hat{g}_q (\omega') \hat{\Sigma}_\phi^{(1)} \partial_\phi \hat{g}_q(\omega' + \Omega)  
\right]^<\\
\nonumber
=&&\frac{iJ^2}{4} \int\frac{dq}{2\pi} \int\frac{d\omega'}{2\pi} {\rm Tr}\left[
\hat{j}_s \hat{g}_q (\omega') \hat{\Sigma}_\phi^{(1)} \hat{g}_q (\omega') \hat{\Sigma}_\phi^{(1)}  \hat{g}_q(\omega' + \Omega) 
\right.\\
\nonumber
&& \left. \hspace{2.5cm}
+ \hat{j}_s \hat{g}_q (\omega') \hat{\Sigma}_\phi^{(1)}  \hat{g}_q(\omega' + \Omega) \hat{\Sigma}_\phi^{(1)} \hat{g}_q(\omega' + \Omega) 
\right]^<\\
=&& -\frac{K_p(\Omega, 0) + K_p(\Omega, \Omega)}{2},
\label{eq:relation}
\eea
where we have used the relationship, $\partial_\phi \hat{g}_q (\omega) = -J \hat{g}_q (\omega) \hat{\Sigma}_\phi^{(1)} \hat{g}_q (\omega)$.
Note that the surface term vanishes as the first-order susceptibility of the spin current is zero.
With the relationship, Eq.~\eqref{eq:relation}, the expansion coefficients of the diamagnetic contributions are described by those of the paramagnetic contributions by
\bea
\chi_d^{(0)} &=& -\chi_p^{(0, 0)},\\
\chi_d^{(1)} &=& - \chi_p^{(1, 0)} - \frac{\chi_p^{(0,1)}}{2},\\
\chi_d^{(2)} &=& - \chi_p^{(2, 0)} - \chi_p^{(1, 1)} - \frac{\chi_p^{(0, 2)}}{2}.
\eea
By substituting the relations into Eq.~\eqref{eq:append_js}, the pumped spin current is finally obtained as
\bea
\braket{j_s}(t) = -\frac{\chi_p^{(0, 2)}}{2} \left(\frac{d\phi (t)}{dt}\right)^2
\eea
where the all gauge-dependent terms of the diamagnetic and paramagnetic contributions are cancelled each other.

\subsubsection{The second-order susceptibility}
Finally, let us derive the second-order response function responsible for the spin current.
The pumped spin current is given by the response function $\chi_p^{(0, 2)}$, which is evaluated as
\begin{widetext}
\bea
\nonumber
\chi_p^{(0, 2)} &=& i2J^2 \int \frac{dq}{2\pi} \int \frac{d\omega'}{2\pi}
\left(\hat{j}_s\right)_{ii} \left(\hat{\Sigma}_\phi^{(1)}\right)_{ij} \left(\hat{\Sigma}_\phi^{(1)}\right)_{ji}\\
&&
\nonumber
\times f(\omega')
\left[
\left\{
\left(g_i^r(\omega') \right)^3 g_i^a(\omega')
+
\left( g_i^r(\omega') \right)^2 \left( g_i^a(\omega') \right)^2
+
g_i^r(\omega') \left( g_i^a(\omega') \right)^3
\right\}
\left\{
g_j^r(\omega') - g_j^a(\omega') 
\right\}
\right.\\
\nonumber
&&\hspace{1cm}
\left.
+
\left\{g_{q, i}^r(\omega')  - g_{q, i}^a(\omega') \right\}
\left\{
g_{q, i}^r(\omega') \left(g_{q, j}^r(\omega') \right)^3 + g_{q, i}^a(\omega')  \left(g_{q, j}^a(\omega') \right)^3
\right\}
\right]\\
\nonumber
&=& -6 J^2 \int \frac{dq}{2\pi} \sum_{i \ne j}\frac{f(\epsilon_i - \EF)}{\left(\epsilon_i - \epsilon_j\right)^4}
\left[ \left(\hat{j}_s\right)_{ii} - \left(\hat{j}_s\right)_{jj} \right] \left|\left(\hat{\Sigma}_\phi^{(1)}\right)_{ji}\right|^2\\
&&
-\frac{i J^2}{4} \int \frac{dq}{2\pi} \sum_{i \ne j} 
\left(\hat{j}_s\right)_{ii}\left|\left(\hat{\Sigma}_\phi^{(1)}\right)_{ji}\right|^2
\left\{ \left( g^r_{q, i} \right)^2 - \left( g^a_{q, i}\right)^2 \right\} \left(g^r_{q, j} + g^a_{q, j}\right)^2,
\label{eq:js_res}
\eea
\end{widetext}
where $g^{r (a)}_{q, i} \equiv g^{r (a)}_{q, i} (0)$ and we have assumed $1 << \EF \tau$ corresponding to a weak impurity scattering regime and retained terms in the leading order of $1/\tau$.
Note that terms containing off-diagonal components of the spin current operator vanishes.
The first term of Eq.~\eqref{eq:js_res} emerges from contributions of all the states below the Fermi energy, known as the Fermi sea contribution, and describes a non-dissipative spin current which is independent of impurity scattering.
On the other hand, the second term of Eq.~\eqref{eq:js_res} contains contributions from excitations around the Fermi surface, therefore, gives dissipative spin current.
Finally, the pumped spin current is obtained as
$
{j_s}(t) = \left(\chi^{(2)}_{ND} + \chi^{(2)}_{D}\right) \left(\frac{d\phi(t)}{dt}\right)^2,
$
where $\chi^{(2)}_{ND (D)}$ is the non-dissipative (dissipative) part of the second-order susceptibility defined by
\bea
\nonumber
\chiND &=& 3 J^2 \int \frac{dq}{2\pi} \sum_{i \ne j}\frac{f(\epsilon_i - E_i)}{\left(\epsilon_i - \epsilon_j\right)^4}\left[ \left(\hat{j}_s\right)_{ii} - \left(\hat{j}_s\right)_{jj} \right] \left|\left(\hat{\Sigma}_\phi^{(1)}\right)_{ji}\right|^2,\\\\
\nonumber
\chiD &=& \frac{i J^2}{8} \int \frac{dq}{2\pi} \sum_{i \ne j} 
\left(\hat{j}_s\right)_{ii}\left|\left(\hat{\Sigma}_\phi^{(1)}\right)_{ji}\right|^2\\
&&\hspace{1.5cm}
\times\left\{ \left( g^r_{q, i} \right)^2 - \left( g^a_{q, i}\right)^2 \right\} \left(g^r_{q, j} + g^a_{q, j}\right)^2.
\eea


\section{Analytical calculations in 2-band continuum model}
\label{sec:continuum}

\begin{figure*}[htbp]
\centering
\includegraphics[width = \textwidth]{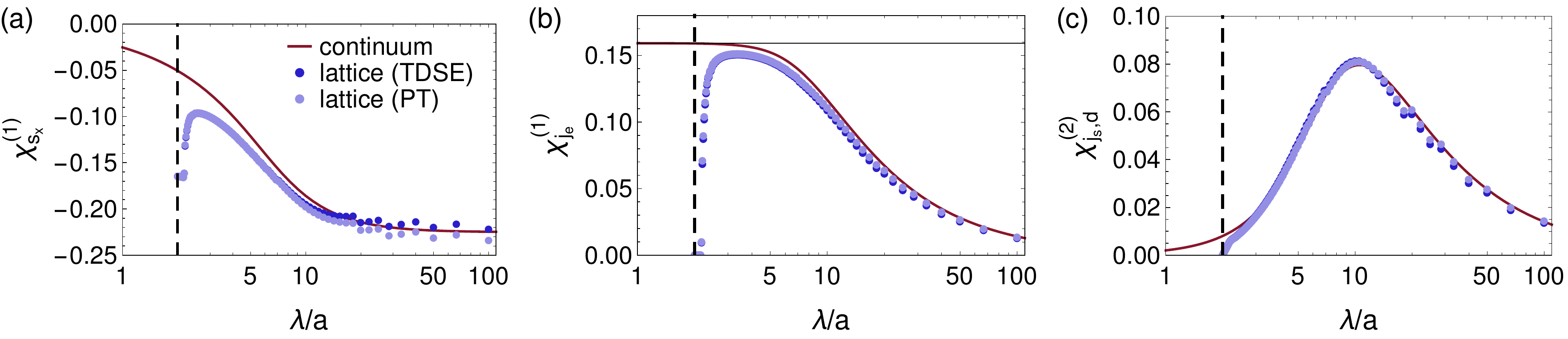}
\caption{
Pitch length dependence of the susceptibilities corresponding to (a) spin polarization, (b) charge current, and (c) non-dissipative contribution of spin current.
The solid lines indicate the results obtained by the continuum model, see Eqs.~\eqref{cont_sx}, \eqref{cont_je}, and \eqref{cont_js}.
Dots are numerical data obtained by the time-dependent Shr\"odinger equation (TDSE) and perturbation theory (PT), respectively.
Parameters are $\omega = 0.3$, $\tH = 1$, $m_e = \tH/2$, $J = 0.5$, $\EF = -1$, and $\tau = 5$.
}
\label{figS1}
\end{figure*}

In this appendix, we analytically derive the susceptibilities in a two-band continuum model.
As a continuum Hamiltonian describing one-dimensional electrons coupled to a spiral spin order, we have considered
\bea
\nonumber
H = \int \frac{dk}{2\pi} \left[
\bm c_k^\dagger \left(
\frac{k^2}{2m_e} - \EF
\right)\bm c_k^{} 
-J
\bm m_Q (t) \bm c_{k+\frac{Q}{2}}^\dagger \bm \sigma \bm c_{k-\frac{Q}{2}}^{}  + \rm h.c.\right]\\
\eea
where $\bm c_k$ is the Fermionic annihilation operator with momentum $k$, $\EF = \frac{Q^2}{8m_e}$ is the Fermi energy, $m_e$ is an effective electron mass, and $\bm m_Q(t) = \frac{\hat{\bm y} + i \hat{\bm z}}{2} e^{-i\phi(t)} \approx \frac{\hat{\bm y} + i \hat{\bm z}}{2}\left[1 - i\phi(t) - \frac{1}{2}\phi^2(t)\right]$ describes the spiral magnetic order with wave vector $Q$.
Here we have assumed that the Fermi energy is located in the SDW gap.
By introducing the spinor, $\Psi_k = \left[c_{k+\frac{Q}{2}, \uparrow}, c_{k-\frac{Q}{2}, \downarrow}\right]$ whose spin quantization axis is taken along the propagation vector $\bm Q$,
the Hamiltonian is rewritten as $H = \int \frac{dk}{2\pi} \Psi_{k}^\dagger \left[ \hat{h}_0(k) + J\phi(t) \hat{\Sigma}^{(1)}_\phi + \frac{J}{2}\phi^2(t) \hat{\Sigma}^{(2)}_\phi\right] \Psi_{k}^{}$, where the bare Hamiltonian is
$
\hat{h}_0(k) = \frac{k^2}{2m_e} \tau_0 + \frac{Qk}{2m_e} \tau_z - J \tau_x,
$
the spin operators coupled to the phason are
$
\hat{\Sigma}_\phi^{(1)} = \tau_y
$
and
$
\hat{\Sigma}_\phi^{(2)} = \tau_x
$
, and $\bm \tau$ is a vector of the Pauli matrix.
The charge and spin current operators are defined by $\hat{j}_e = -e\left(\frac{k}{m_e} \tau_0 + \frac{Q}{2m_e} \tau_z \right)$ and $\hat{j}_s = -e\left(\frac{k}{m_e} \tau_z + \frac{Q}{2m_e} \tau_0 \right)$, respectively.
The eigenvalues of $\hat{h}_0(k)$ are given by $\epsilon_{k,\pm} = \frac{k^2}{2m_e} \pm \sqrt{\left(\frac{kQ}{2m_e}\right)^2 + J^2}$.
From Eqs.~(\ref{eq:order1}),~(\ref{eq:order2nd}), and~(\ref{eq:order2d}), the susceptibilities with the continuum model are obtained as
\bea
\chi_{s_x}^{(1)} &=& -\frac{\sqrt{m_e}}{\pi} \left(\zeta + \frac{Q^2}{2m_e}\right)^{-\frac{1}{2}},
\label{cont_sx}\\
\chi_{j_e}^{(1)} &=& \frac{eQ}{2\pi \sqrt{m_e}} \left(\zeta + \frac{Q^2}{2m_e}\right)^{-\frac{1}{2}},
\label{cont_je}\\
\chiND &=& \frac{eQ}{2\pi \sqrt{m_e}} \left(\zeta + \frac{Q^2}{2 m_e}\right)^{-\frac{3}{2}},
\label{cont_js}
\eea
where $\zeta = \sqrt{\left(\frac{Q^2}{2m_e}\right)^2 + 4 J^2}$.

For the short-pitch spiral, $\lambda < \pi/(2\sqrt{m_e J})$, each susceptibility is given by
$
\chi_{s_z}^{(1)} = -m_e \lambda/(2\pi^2)
$,
$
\chi_{j_e}^{(1)} =  e/(2\pi)
$, and
$
\chiND = e m_e \lambda^2/(8\pi^3).
$
In this regime corresponding the adiabatic regime, the Thouless pumping is realized as the charge current susceptibility is quantized.
On the other hand, for the long-pitch spiral where $\lambda < \pi/(2\sqrt{m_e J})$, the susceptibilities are given by
$
\chi_{s_z}^{(1)} = -\sqrt{m_e/(2\pi^2 J)}
$,
$
\chi_{j_e}^{(1)} =e/(\lambda \sqrt{2 J m_e})
$, and
$
\chiND = e/(2\sqrt{2} \lambda \sqrt{m_e J^3} ).
$
This regime corresponds to the the nonadiabatic regime, and the charge and spin current decay as $\lambda^{-1}$ due to the presence of metallic Fermi surfaces.
Note that, in ferromagnetic limit, $\lambda \rightarrow \infty$, both charge and spin current asymptotically becomes zero.
The pitch-length dependence of the susceptibilities are depicted in Fig.~\ref{figS1} with the results of the lattice model.
In the long-pitch regime, results obtained for the continuum and lattice models agree well.
In contrast, as the pitch length reaches $\lambda/a = 2$, two results show deviations as the effect of discretization becomes important.

\section{Comparison between collinear SDW and spiral}
\label{sec:collinearSDW}

\begin{figure*}[tbp]
\centering
\includegraphics[width = \textwidth]{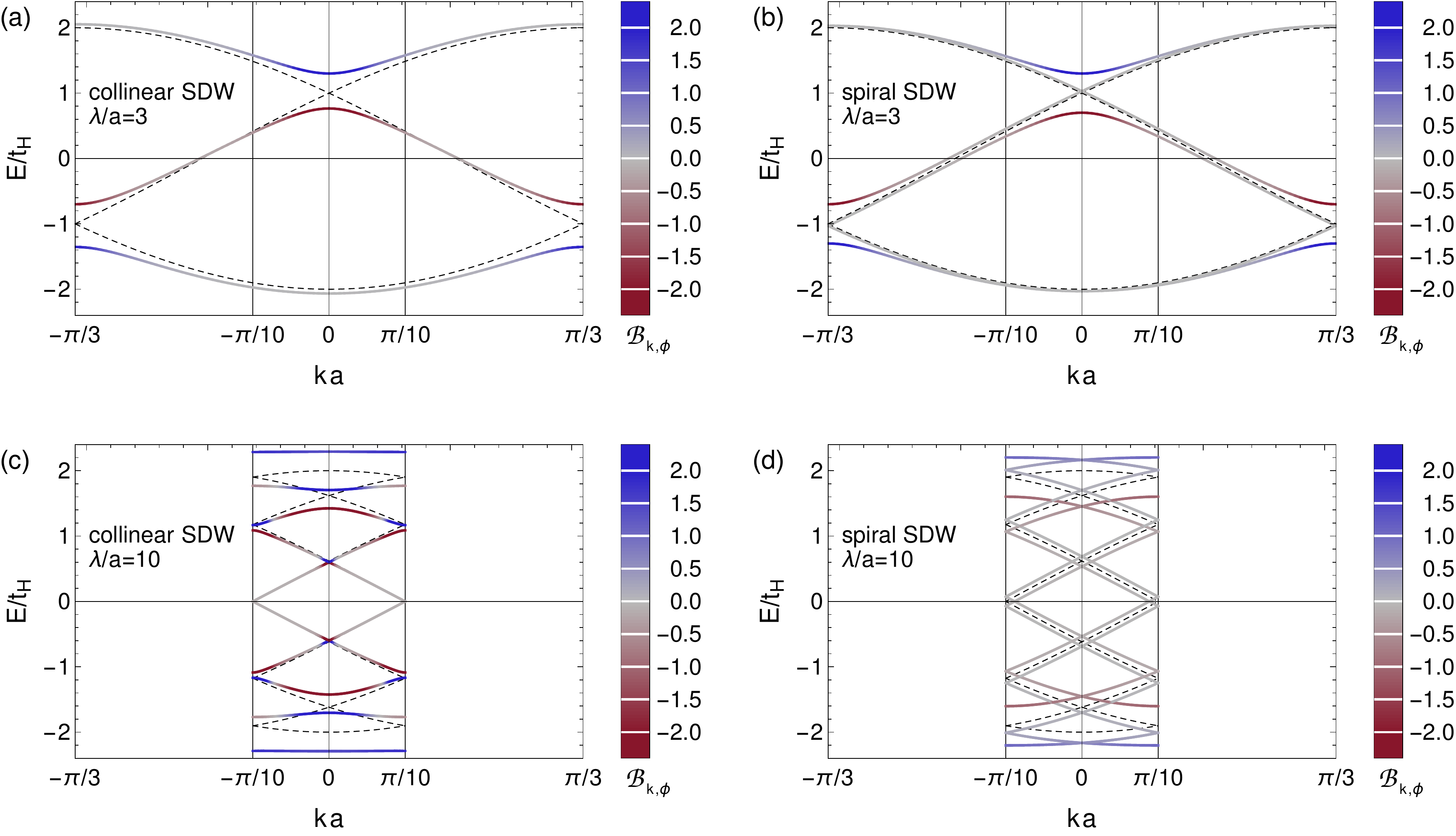}
\caption{
   The dispersion $E(k)$ of a periodic system with $J/\tH = 0.3$.
   Comparison of spin density waves (SDW) with (a,c) collinear order or (b,d) spiral order.
   The wavelength is (a,b) $\lambda/a = 3$ or (c,d) $\lambda/a = 10$, as indicated in the panels. 
   The color encodes the momentum-phason Berry curvature $\mathcal{B}_{k,\phi}$, defined in Eq.~\eqref{eq:Berry}, which is truncated to the same range in every panel, see legends.
}
\label{figS2}
\end{figure*}

\begin{figure}[tbp]
\centering
\includegraphics[width = \columnwidth]{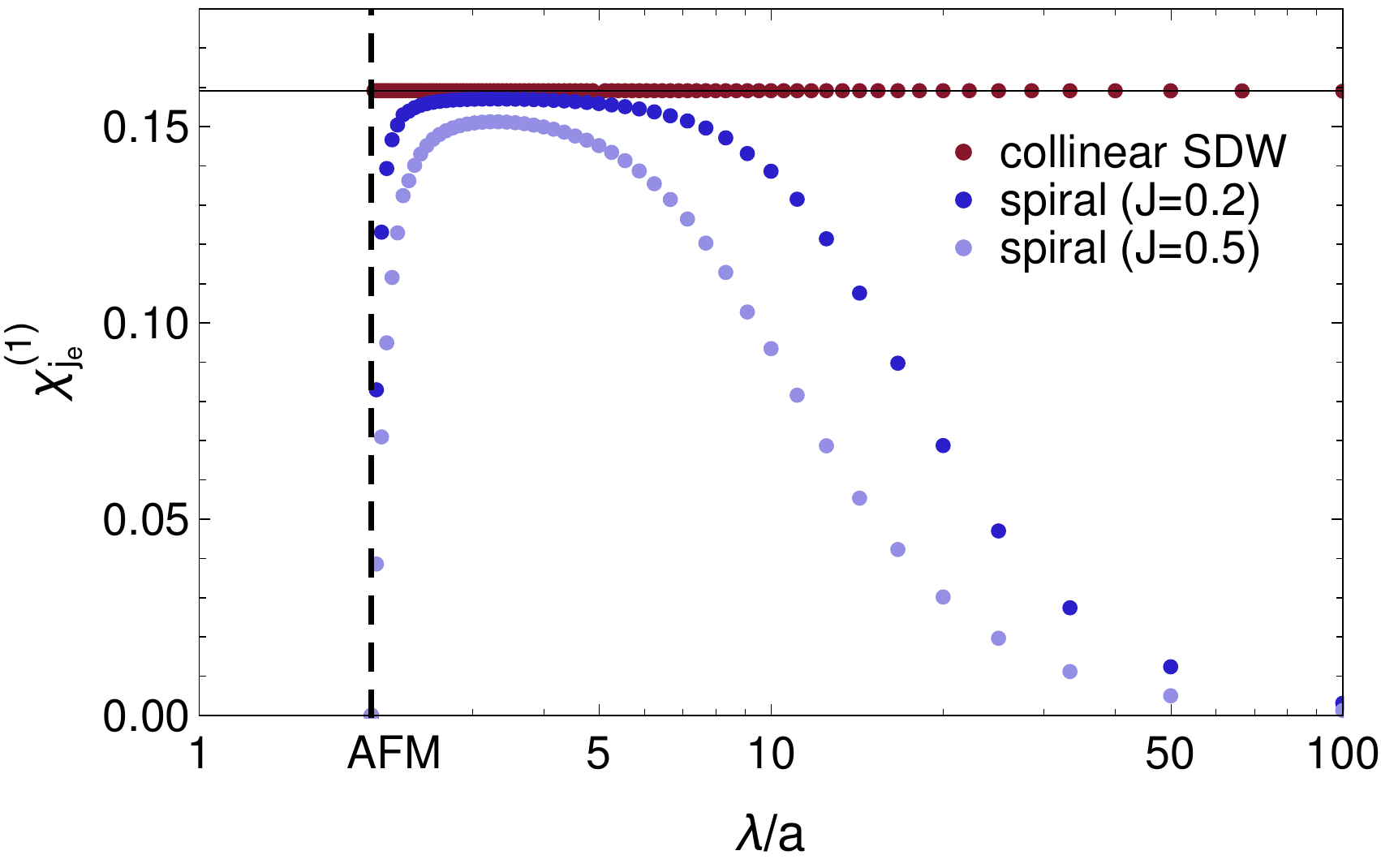}
\caption{
    Dependence of the charge current susceptibility $\chi_{j_e}^{(1)}$ on the pitch length $\lambda$.
    Comparison of results for the collinear and spiral SDW.
    Parameters are $\tH = 1$ and $J=0.2$ or $J=0.5$ as indicated.
    The result for the collinear SDW is quantized, irrespective of the precise value of $J$.
    The Fermi energy $\EF/\tH = -2 \cos(\pi a/\lambda)$ is chosen in the center of the SDW gap.
}
\label{figS2b}
\end{figure}

In the adiabatic Thouless pumping regime, electrons coherently move with the collective coordinate; namely, charge current is expected to be given as $\braket{j_e} = -e n_e v$, where $n_e$ is charge density and $v$ is the drift velocity of the collective coordinate.
One can notice that the charge density is inversely proportional to the pitch length $\lambda$ when the nesting condition is satisfied, while the drift velocity is proportional to $\lambda$ as it is given by $v = -\eta \lambda \omega/(2\pi)$.
Therefore, the pumped charge current as a function of rotation frequency $\omega$ should be independent of $\lambda$.
Contrary, in the ferromagnetic states which corresponds to the long-pitch limit of spiral, $\lambda \rightarrow \infty$, the charge and spin current should vanish.
These two regimes are connected by the crossover between the adiabatic and nonadiabatic regime, as discussed in Sec.~\ref{sec:continuum}.
In this section, we will discuss the crossover in terms of the Berry curvature by including a comparison to the collinear SDW state, see Fig.~\ref{figS2b}.

Magnetization in the collinear SDW state is described by
\begin{equation}
\bm M(x,t) = \bm e_1 \cos \theta(x,t)
\end{equation}
where $\theta(x,t) = Qx + \phi(t)$, and we take polarization along $x$-axis.
In sharp contrast to that of spiral SDW states, an electronic band dispersion of collinear SDW is fully gapped, see Fig.~\ref{figS2} (a) for a band structure with $\lambda/a = 3$.
Note that each band has 2-fold degeneracy, and the degenerate bands have exactly the same Berry curvature.
As shown in Figs.~\ref{figS2} (a) and (b), the Berry curvature is mostly concentrated around the SDW gaps, which resembles to that of the spiral SDW state.

The Berry curvature distribution is extended in the momentum space, and its width is proportional to the exchange constant and the pitch length, as the half width of the Berry curvature $q_w$ is given by $q_w \approx J \lambda \sqrt{2^{3/2} - 1}/(2 \pi a^2 \tH)$, while the maximum value of the Berry curvature is obtained as $\pm\pi a^2 \tH /(J \lambda)$.
When the width $q_w$ is smaller than a size of the first Brillouin zone, the Berry curvature is well localized in the momentum space as shown in Figs.~\ref{figS2} (a) and (b).
In this case, the momentum integral of the Berry curvature for the lower two bands is quantized for both the collinear and spiral SDW states, corresponding to the adiabatic pumping regime.

On the other hand, differences appear when $q_w$ becomes larger than the Brillouin zone.
As shown in Fig.~\ref{figS2} (c) and (d), the Berry curvature density increases as the first Brillouin zone becomes smaller because of folding for the collinear SDW state, while it decreases for the spiral SDW state.
For the collinear SDW state (Fig.~\ref{figS2} (c)), the total Berry curvature of the lower two bands remains invariant because the band structure is fully gapped, and the Berry curvature flux cannot escape.
Recalling that the charge current is proportional to the total Berry curvature of lower two bands, the charge current for the collinear SDW state is quantized to $-e/(2\pi)$ and independent of the pitch-length.

In contrast, for the spiral SDW state (Fig.~\ref{figS2} (d)), the Berry curvature density spreads into the higher bands through the gapless helical channels, and the total Berry curvature of lower two bands decreases.
In this regime of the spiral SDW state, the adiabatic transport picture is no longer applicable as the gapless states also carry the Berry curvature.
Consequently, the quantized transport is only observed for the short-pitch spirals whose pitch length $\lambda < \pi a\sqrt{\tH/(2 J)}$, and the charge current decreases as the pitch-length becomes longer because of the nonadiabaticity.


\section{Extension to the higher dimensions} 
\label{sec:higherdimensions}

\begin{figure*}[htbp]
\centering
\includegraphics[width = \textwidth]{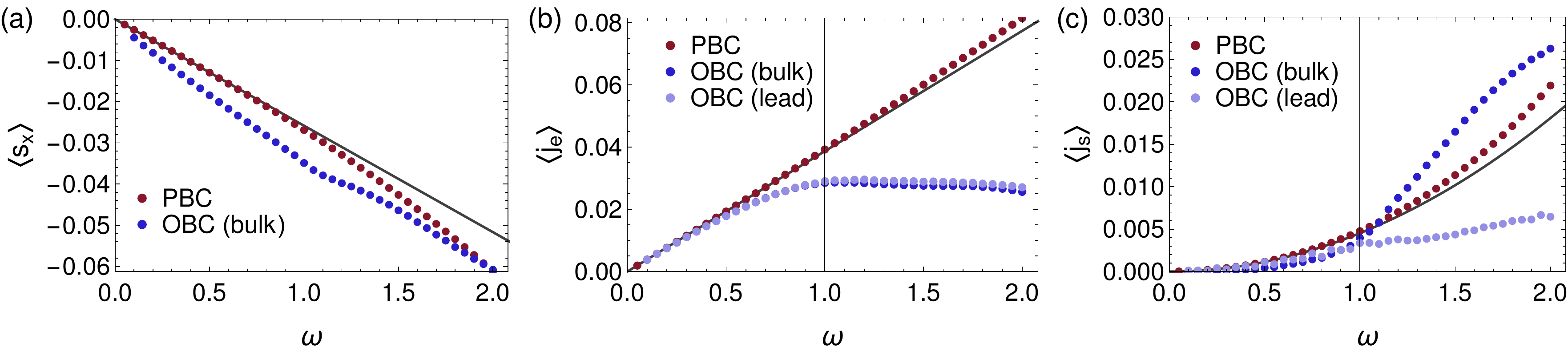}
\caption{
   Averaged (a) spin polarization $\langle s_x \rangle$, (b) charge current $\langle j_e \rangle$, and (c) spin current $\langle j_s \rangle$ as function of the rotation frequency $\omega$ for a two-dimensional system.
   Parameters for all panels are $\lambda/a=3$, $t_\text{\tiny{H}} = 1$, $J = 0.5$, $\tau = 2.5$, and $\EF = -1$.
   }
\label{figS3}
\end{figure*}

We have considered the one-dimensional electronic system for simplicity.
In this section, we will show that the extension to higher-dimension is rather straightforward.
As the spiral order only hybridizes the states along the propagation vector $\bm Q$, the Hamiltonian given in Eq.\eqref{eq:H} is decoupled to terms including perpendicular momenta $\bm k_\perp$; thus, the effect of perpendicular hopping is just to shift the Fermi energy as $\EF \rightarrow \EF - 2\tH \sum_i\cos k_{\perp,i}$.
Similarly, the spin operator, charge and spin current operators along $\bm Q$, $\hat{\Sigma}_{\phi}^{(1)}$, and $\hat{\Sigma}_{\phi}^{(2)}$ are all independent of $\bm k_{\perp}$.
Therefore, the transport response functions for the higher dimensions are simply obtained by integrating the one-dimensional susceptibility over the perpendicular momenta,
$
\chi_{\hat{O}} (\EF) = \sum_{\bm k_\perp} \chi^{1D}_{\hat{O}} (\EF - 2\tH \cos k_{\perp, i}),
$
where $\chi^{1D}_{\hat{O}}$ is the one-dimensional susceptibilities given in Eqs.~\eqref{eq:order1},~\eqref{eq:order2nd}, and~\eqref{eq:order2d}.
This consideration suggests that the pumping effects in the higher-dimensional systems are qualitatively the same as those of one-dimensions.
Note that the {\it almost} universal quantization of the charge current is only present in the one-dimension, and the charge current depends on the other parameters such as the exchange constant $J$ in the higher dimensions.

We further conducted the numerical calculations for a two-dimensional electronic system based on the perturbation theory and the time-dependent Shr\"odinger equations.
As shown in Fig.~\ref{figS3}, spin polarization and charge current show linear dependence of the rotational frequency $\omega$, while spin current is proportional to $\omega^2$.
When the frequency is larger than the SDW gap, $|\omega| < 2|J|$, results obtained for the PBC and OBC coincide.
In contrast, in the higher frequency regime, $|\omega| < 2|J|$, the PBC and OBC results show deviation as the PBC results continue to increase, while the OBC results show saturation.
These behaviors are qualitatively the same with those of the one-dimensional system presented in Fig.~\ref{fig3} (a-c).


\end{document}